\documentclass[%
 reprint,
 amsmath,amssymb,
 aps,
]{revtex4-2}

\usepackage{graphicx}
\usepackage{dcolumn}
\usepackage{bm}

\usepackage{hyperref}
\usepackage{cleveref}
\usepackage {graphicx}
\usepackage{braket}
\usepackage{titlesec}
\usepackage{xcolor}
\usepackage{ulem}
\usepackage{xfrac}

\begin{document}

\preprint{APS/123-QED}

\title{Anomalous Spontaneous Emission Enhancement by Non-Hermitian Momentum-Space Bound States in the Continuum}

\author{Keren Wang}
\altaffiliation{These authors contributed equally to this work}
\affiliation
{College of Physics, Sichuan University, Chengdu 610064, China}

\author{Jing Du}
\altaffiliation{These authors contributed equally to this work}
\affiliation
{College of Physics, Sichuan University, Chengdu 610064, China}

\author{Wei Wang}
\email{w.wang@scu.edu.cn}
\affiliation
{College of Physics, Sichuan University, Chengdu 610064, China}

\date{\today}

\begin{abstract}

    Conventional Purcell theory emphasizes high quality factors ($Q$) for spontaneous emission (SE) enhancement in cavities, but overlooks collective Bloch mode effects in periodic nanostructures like photonic crystal slabs. We introduce a unified temporal coupled-mode framework to evaluate Purcell and photoluminescence factors through momentum-space integration, revealing anomalous SE enhancement by non-Hermitian momentum-space bound states in the continuum (BICs). In periodic dielectric structures with comparable effective mode volumes, this yields substantial SE enhancement in low-$Q$ regimes—defying the traditional high-$Q$ paradigm and inversely correlated with system $Q$—while emission rates are stably twice the photoluminescence, eliminating critical coupling requirements. Unique spectral profiles, contradicting Lorentzian/Fano assumptions, arise from collective mode interactions. Full-wave simulations confirm these challenges to conventional wisdom, with non-Hermitian BICs outperforming high-$Q$ designs across broad numerical apertures. This establishes a novel paradigm leveraging non-Hermiticity and topological protection for robust, bright emitters, redefining nanophotonic applications in lasers and light-emitting diodes.
\end{abstract}

\maketitle

The enhancement of spontaneous emission (SE) in optical cavities has long been a cornerstone of photonics. Traditionally, the Purcell factor establishes a direct connection between the cavity quality factor ($Q$) and the emission rate\cite{PhysRev.69.674}, motivating decades of efforts toward high-$Q$ resonators\cite{RN968,RN1253,RN1235}, such as plasmonic cavities\cite{RN1237, RN1253} and photonic crystal defect modes\cite{RN1233,RN937,RN1255}.

More recently, periodic nanostructures—such as photonic crystal slabs\cite{Joannopoulos2008PhotonicCM}—have attracted tremendous attention. These systems can support resonances with arbitrarily high $Q$ factors through bound states in the continuum (BICs) and quasi-guided modes (QGMs)\cite{RN296,RN475,RN872}. While such resonances hold great promise for emission enhancement, a crucial yet often overlooked feature of periodic systems is their inherent support of an infinite continuum of orthogonal Bloch modes, each associated with a distinct in-plane wavevector ($k$). Unlike isolated resonators, where a dipole emitter couples predominantly to a single mode\cite{RN1234,RN914}, an emitter embedded in a periodic structure excites multiple Bloch modes across the entire momentum space at a given frequency. This leads to a collective contribution that fundamentally alters the emission process.

This distinction is central to two complementary quantities: \textcolor{black}{the local density of states (LDOS)\cite{RN1243,RN1011}, which accounts for the total emission power, giving rise to the Purcell factor via comparison with the free-space emission rate}, and the photoluminescence (PL) factor\cite{RN975,RN967,RN1247,RN988,RN1221,RN1246,RN1229}, which reflects the radiation energy collected in the far field. Two theoretical frameworks have been developed in parallel—LDOS based on Fermi’s golden rule\cite{RN1234, RN995,RN1007}, and PL based on local field enhancement\cite{RN1004,RN1220,RN746}, linked through Lorentz reciprocity\cite{RN1223}. Yet, existing theories often consider only specific wavevectors\cite{RN1245,RN1250} and neglect the collective role of Bloch modes, leaving an incomplete picture of light–matter interactions in periodic structures. A unified framework specialized for periodic systems is still lacking.

Here, we demonstrate that this collective effect yields phenomena that challenge conventional intuition. We first develop a generalized temporal coupled-mode theory (TCMT)\cite{RN1249,RN239}, grounded in input-output theory\cite{PhysRevA.31.3761}, for multi-mode, non-Hermitian systems. This framework enables simultaneous evaluation of the Purcell factor ($F_p$) and the photoluminescence enhancement factor ($F_\text{PL}$). Applying this framework to a minimal yet common two-level system, we quantify the collective effect by integrating emission power across momentum space. \textcolor{black}{Surprisingly, we find an anomalous SE enhancement near BICs, which is negatively correlated with the Q factors at every wavevector, contradicting the conventional Purcell formula.} We validate our prediction through grating designs that achieve substantial SE enhancement with regular mode volume, unattainable with conventional design approaches. Furthermore, we find that the total emission power is consistently twice the PL power, i.e., $F_p \approx 2 F_\text{PL}$, indicating the absence of critical coupling between radiative and non-radiative channels, which relaxes fabrication tolerances and enhances design flexibility. The frequency-domain lineshape is also unique, exhibiting neither Lorentzian nor Fano characteristics. Finally, our framework encompasses both BICs and QGMs, providing precise guidance for designing and optimizing high-performance nanophotonic emitters in versatile periodic systems.

\begin{figure}
    \centering
    \includegraphics[width=1.0\linewidth]{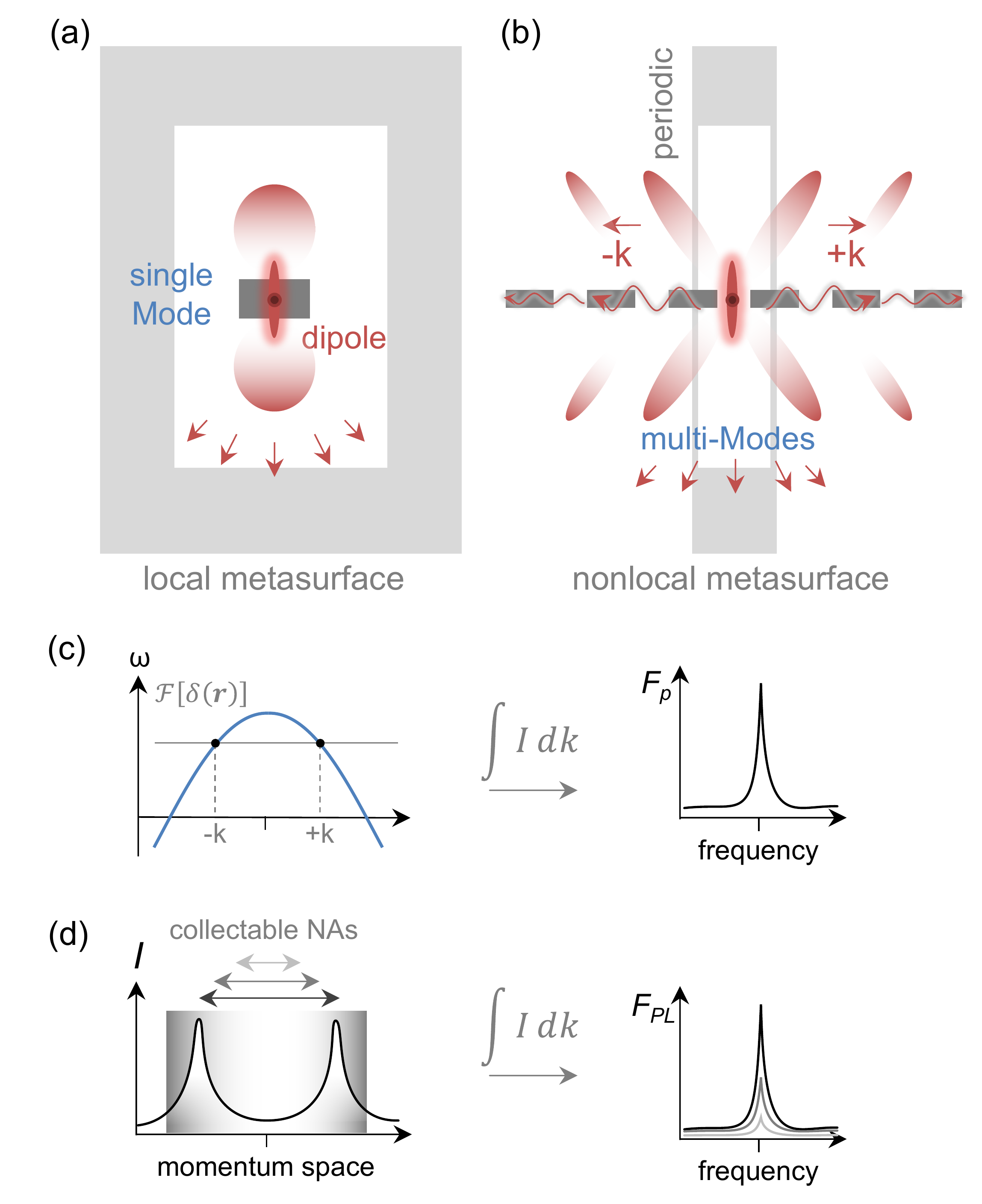}
    \caption{Schematic of SE enhancement in local and nonlocal metasurfaces. (a) Local metasurface with absorbing boundaries (e.g., PML), where dipoles couple to discrete modes. (b) Nonlocal metasurface with periodic boundaries, supporting orthogonal Bloch modes labeled by in-plane wavevector $\mathbf{k}$. (c) Continuous band structure in momentum space, where a dipole excites all modes at a given frequency. (d) PL collected from the radiative power $P_{\rm rad}$ within a limited NA in momentum space.}
    \label{fig:principle}
\end{figure}

As shown in \autoref{fig:principle}, SE enhancement differs fundamentally between local and nonlocal \textcolor{black}{systems}.
\textcolor{black}{
In local systems with open boundaries (\autoref{fig:principle}(a))\cite{RN1234}, a dipole couples to modes\cite{RN1254,RN1234} characterized by discrete order $n$. In contrast, the periodic Bloch boundary conditions introduce the in-plane wavevector $\mathbf{k}$ as another continuous label, thereby enabling periodic metasurfaces (\autoref{fig:principle}(b)) to support orthogonal Bloch modes across momentum space (\autoref{fig:principle}(c))\cite{RN1243} and allowing a dipole at fixed frequency to excite infinitely many k-states.\cite{RN1277}
}
Consequently, the LDOS, or Purcell factor ($F_p$), is evaluated by integrating the total emission power $P_{\rm tot}$ over the entire momentum space (\autoref{fig:principle}(c)). The PL factor ($F_\text{PL}$), however, is determined by integrating the far-field radiation power $P_{\rm rad}$ within a restricted momentum space limited by the numerical aperture (NA) of the objective lens (\autoref{fig:principle}(d))\cite{RN1250}.

Before modeling the collective effects, we first developed a general framework, which separates the total emitted power into radiative and absorptive parts: $P_{\rm tot}=P_{\rm rad}+P_{\rm abs}$ which is needed to evaluate Purcell factor $F_p$ and PL factor $F_\text{PL}$.
We model $m$ resonant modes using an effective non-Hermitian Hamiltonian $H_{\rm eff}=H_0-\frac{i}{2}\Gamma_{\rm abs}-\frac{i}{2}KK^\dagger$, where $H_0=H_0^\dagger\in\mathbb{C}^{m\times m}$ is the Hermitian part, $\Gamma_{\rm abs}\succeq 0$ is the absorption matrix, and $K\in\mathbb{C}^{m\times n}$ is the coupling matrix with $n$ radiation channels. The system response at frequency $\omega$ is governed by the Green’s function $G(\omega) = [\omega I - H_{\rm eff}]^{-1}$. Under input-output theory\cite{PhysRevA.31.3761}, the emission power under dipole excitation, represented by the vector $\mathbf{d} \in \mathbb{C}^m$ (where $\mathbf{d}$ denotes the complex amplitudes of the dipole moments exciting the $m$ resonant modes), is decomposed as follows:
\begin{align}
P_{\rm rad}(\omega)&\propto \mathbf d^\dagger G^\dagger(\omega)\,KK^\dagger\,G(\omega)\,\mathbf d,
\label{eq:Prad_final}\\
P_{\rm abs}(\omega)&\propto \mathbf d^\dagger G^\dagger(\omega)\,\Gamma_{\rm abs}\,G(\omega)\,\mathbf d,
\label{eq:Pabs_final}\\
P_{\rm tot}(\omega)&\propto -\,\mathbf d^\dagger \Im[G(\omega)]\,\mathbf d,
\label{eq:Ptot_final}
\end{align}
Physically, Eq.~\eqref{eq:Prad_final} isolates far-field PL, Eq.~\eqref{eq:Pabs_final} captures non-radiative loss, and Eq.~\eqref{eq:Ptot_final} recovers the conventional LDOS\cite{RN775,RN913,RN709}. Their ratio, $\eta(\omega)=\frac{P_{\rm rad}(\omega)}{P_{\rm tot}(\omega)}$, gives the radiation efficiency, equivalent to the external quantum efficiency (EQE)\cite{RN1257}. Full derivations, along with an alternative perspective using QNMs, are provided in Supplementary Materials Sec. 1\cite{supp1}. This compact formulation enables clear decomposition of energy flows. When absorption- and radiation-related losses are comparable, this leads to phenomena such as critical coupling\cite{RN1220}. Particularly when interacting with ultrahigh-$Q$ resonances (e.g., BICs), this separation is essential, even in low-loss dielectrics\cite{RN250}. We note that similar energy-partition treatments have recently been discussed for gain–loss systems\cite{RN1007}; in contrast, our formulation targets passive structures and PL observables, providing a directly applicable tool for nanophotonic emitters.

We proceed in two steps. First, we validate our power-separation theory using a representative grating system. By analyzing single-wavevector properties, we recover established results, such as the BIC dark state and critical coupling for maximal PL in QGMs. Second, we propose an integration model that reveals collective effects unique to periodic structures, bringing several unexpected properties that are ultimately confirmed by full-wave simulations.

\begin{figure}
    \centering
    \includegraphics[width=1.0\linewidth]{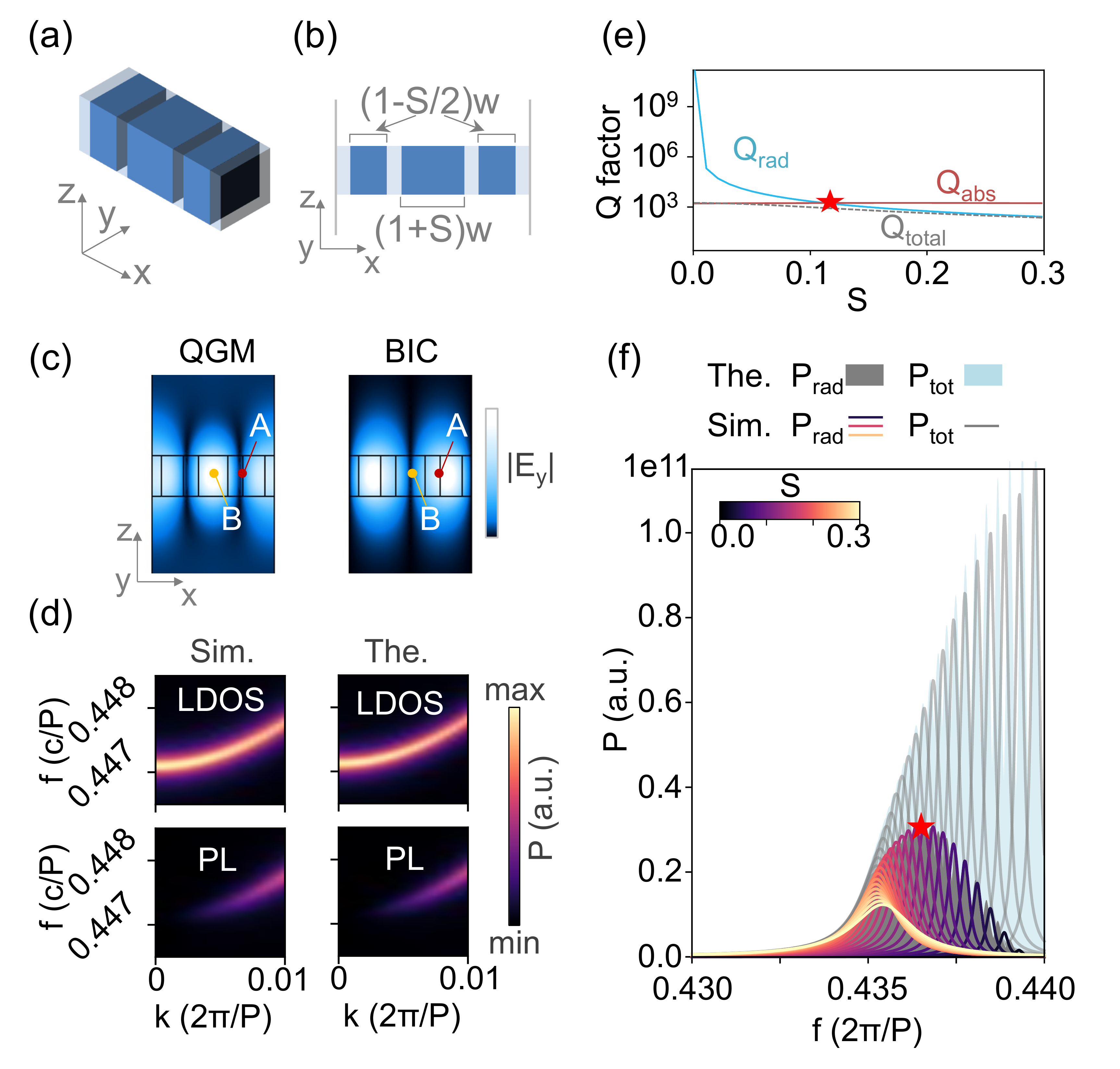}
    \caption{
    (a) Schematic of the ternary folded grating. (b) Grating structural parameters. (c) TE mode field profiles of QGM and BIC at the $\Gamma$ point. (d) Radiation spectra under BIC excitation near the eigenfrequency of BIC: simulations (left) and analytical model (right), comparing LDOS ($P_\text{tot}$, up) and PL ($P_\text{rad}$, bottom). (e) Mode $Q$ factor versus folding strength $S$. (f) QGM excitation: total emission (gray) and PL (colored) under different folding strength $S$ in consistent with (e). The colored background represents analytical model.  
    }
    \label{fig:singleMode}
\end{figure}

We implement a ternary Brillouin-zone folding (BZF)\cite{RN245} in a silicon grating (\autoref{fig:singleMode}(a, b)) to construct a well-defined pair of orthogonal TE-polarized BIC and QGM modes at the $\Gamma$ point, with their field profiles shown in \autoref{fig:singleMode}(c).
The ternary BZF is realized by periodically modulating the widths of every three grating ridges by the factor $S$, while maintaining a constant filling factor ($\text{DC} = 0.6$) and thickness ($t = a/3$, where $a$ is the lattice constant), as illustrated in \autoref{fig:singleMode}(b). For comparison, binary folding produces different effects, as discussed in the Supplementary Materials Sec. 5.\cite{supp1}
The silicon grating has a refractive index of 3.5, and to capture realistic non-radiative losses, we introduce an imaginary part of $10^{-3}$ to the refractive index. By varying $S$, we effectively control the system's non-Hermiticity: the symmetry-protected BIC exhibits vanishing radiation loss, while the QGM acquires radiative decay $\gamma \propto S^2$ (\autoref{fig:singleMode}(e)) and the detuning between modes scales as $\delta \propto S$ (see Supplementary Materials Sec. 5 for details\cite{supp1})\cite{RN220}. 

Within perturbation theory, the effective Hamiltonian of this open system can be written as\cite{RN913,RN384}
\begin{equation}
H_{\rm eff}(k) = 
\begin{pmatrix}
\omega_0+\delta-i \gamma_0 & vk \\
vk &\omega_0 -\delta-i (\gamma_0 + \gamma)
\end{pmatrix},
\label{eq:Heff}
\end{equation}
where $\omega_0$ denotes the central frequency, $\delta$ the frequency detuning at the $\Gamma$ point, $k$ the in-plane wavevector, and $v$ the dispersion strength in momentum space. Here, $\gamma$ corresponds to the radiative loss of the QGM at the $\Gamma$ point, while non-zero $\gamma_0$ represents the overall absorption loss. For simplicity, we neglect differences in absorption losses between the two modes.
Using this effective Hamiltonian, we calculate SE spectra according to Eq.~\eqref{eq:Prad_final}--Eq.~\eqref{eq:Ptot_final} under BIC or QGM excitation. In simulation, as shown in \autoref{fig:singleMode}(c), dipoles placed at locations A or B enable selective excitation of BIC or QGM, and the results are in good agreement with our theory. Near the $\Gamma$ point, LDOS and PL differ strongly: BIC excitation yields no far-field enhancement (dark state in \autoref{fig:singleMode}(d)), whereas QGM excitation produces strong PL, maximized under critical coupling when $S\approx0.11$ (\autoref{fig:singleMode}(e,f)). Details of the theoretical calculations are provided in the Supplementary Materials Sec. 5.\cite{supp1}

With the validity of our theory for computing $P_{\rm tot}$ and $P_{\rm rad}$ confirmed through single-wavevector analysis, we now extend it to capture the collective effects across momentum space.
To quantify these collective effects, we integrate over the momentum space to obtain the Purcell factor $F_p(\omega)$ and the PL factor $F_{\rm PL}(\omega)$:
\begin{align}
F_p(\omega) &\propto \int P_{\rm tot}(\omega,k)\, dk, \label{eq:Fp_int}\\
F_{\rm PL}(\omega) &\propto \int P_{\rm rad}(\omega,k)\, dk. \label{eq:FPL_int}
\end{align}
where the integrals sum over all Bloch modes in momentum space.
Three approximations are implicit here. 
\textcolor{black}{
First, for a non-ideal point source emitter, the finite size limits the effective coupling to high-$k$ modes, thus imposing finite integration bounds.
}
Second, PL collection optics have a limited NA, imposing a cutoff $|k| < {\rm NA} \cdot k_0$, where $k_0$ is the free-space wavevector. More rigorous consideration is to set $F_{\rm PL}(\omega) \propto{ \int_{-{\rm NA} \cdot k_0}^{{\rm NA} \cdot k_0} P_{\rm rad}(\omega, k)\, dk}/{(2{\rm NA} \cdot k_0)}$, as elaborated in Supplementary Materials Sec. 2.\cite{supp1}
Third, far-field measurements are performed in angular coordinates, requiring $d\theta$ rather than $dk$.Nevertheless, this approach adequately captures the collective effects and aligns well with full-wave simulations, so we proceed without these refinements here. Further details on these approximations are provided in Supplementary Materials Sec. 2.\cite{supp1}

For $\delta=0$ in Eq.~\eqref{eq:Heff}, two exceptional points (EPs) emerge at $k_{\rm EP}^{\pm}=\pm\gamma/(2v)$, leading to a flat dispersion with width $\Delta k_{\rm flat}=\gamma/v$. We term this a non-Hermiticity-induced flat band, which remains approximately valid for large $\gamma \gg \delta$, as discussed in Supplementary Materials Sec. 3.\cite{supp1} In this regime, the analytical estimates of $F_p$ and $F_{\rm PL}$ for the BIC are (see Supplementary Materials Sec. 2 for derivations\cite{supp1}):
\begin{align}
F_p^{\rm BIC}(\omega=0) &\propto \frac{2\pi(\gamma_0+\gamma)}{\sqrt{\gamma_0(\gamma_0+\gamma)}}, \label{eq:Fp_BIC}\\
F_{\rm PL}^{\rm BIC}(\omega=0) &\propto \frac{\pi\gamma}{\sqrt{\gamma_0(\gamma_0+\gamma)}}. \label{eq:FPL_BIC}
\end{align}
Both factors increase monotonically with $\gamma$, defying conventional expectations that higher $Q$ factors yield stronger emission. Moreover, the frequency domain lineshape will be unique from any Lorentz or Fano lineshape\cite{RN1234}. This atypical profile originates from the collective overlap of multiple Bloch modes across momentum space, rather than from non-Hermiticity alone\cite{RN1233}.  Surprisingly, in the limit $\gamma \gg \gamma_0$, these expressions simplify to
\begin{equation}
\frac{F_p^{\rm BIC}}{F_{\rm PL}^{\rm BIC}}
=2\Bigl(1+\frac{\gamma_0}{\gamma}\Bigr)
\xrightarrow[\gamma\gg\gamma_0]{} 2,
\label{eq:twice_relation_Fp_FPL}
\end{equation}
indicating the absence of critical coupling between the radiation and absorption channels. Consequently, the radiation efficiency approaches its maximum of 0.5 with small absorptions. Finally, for QGM excitation under the same $\delta \approx 0$ condition, the analysis predicts another counterintuitive outcome, where the Purcell factor $F_p^{\rm QGM}(\omega)$ shows a dip, rather than a peak, at the central frequency. The theoretical spectra are given in Supplementary Materials Sec. 2 and 3.\cite{supp1}

\begin{figure}
    \centering
    \includegraphics[width=1.0\linewidth]{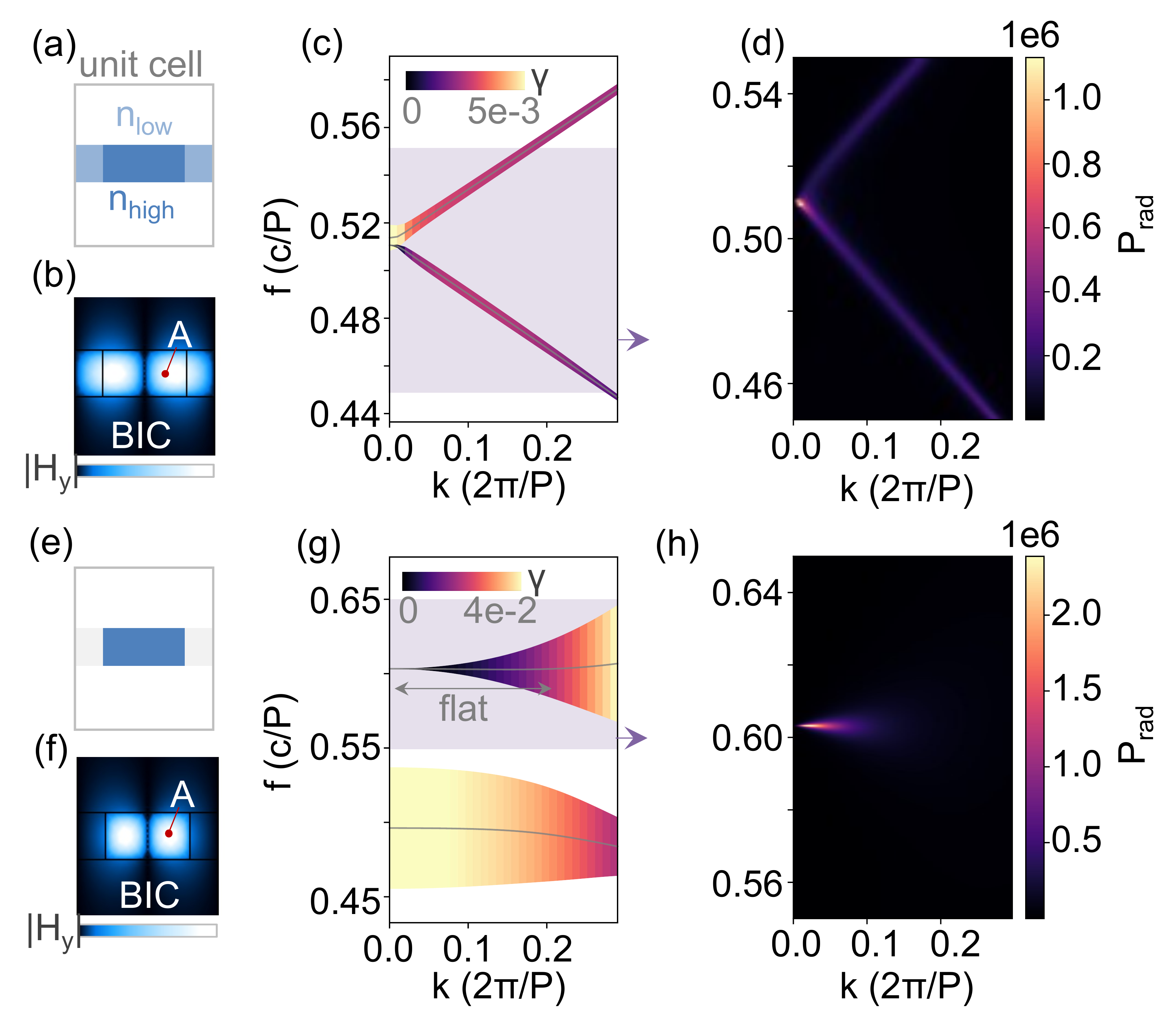}
    \caption{(a) Grating with weak refractive index contrast. (b) Supported TM-BIC mode magnetic field distribution. (c) Momentum-space band structure, with linewidth and color determined by the imaginary part of eigenfrequencies. (d) Momentum-space far-field PL spectra. (e-h) Results for a silicon grating with strong non-Hermiticity, exhibiting low radiation $Q$ factors yet retaining a flat-BIC.}
    \label{fig:TM_band}
\end{figure}

Building on these theoretical predictions, we validate them using full-wave simulations. As indicated in \autoref{fig:TM_band}(a, e), we model a one-dimensional silicon grating with refractive indices $n_\text{high} = 3.5 + 10^{-3}i$ and $n_\text{low} = \Re[n_\text{low}] + 10^{-3}i$, analyzing the behavior of TM modes (\autoref{fig:TM_band}(b, f)). The grating maintains a constant filling factor ($\text{DC} = 0.6$) and thickness ($t = a/3$). The effective Hamiltonian in Eq.~\eqref{eq:Heff} remains sufficiently applicable to this system. By tuning the refractive index contrast ($\Re[n_\text{high}] - \Re[n_\text{low}]$), we control the system’s non-Hermiticity. In the strongly non-Hermitian regime ($\Re[n_\text{low}]=1$), the BIC remains robust despite a sharp decrease in its $Q$ factor under symmetry breaking, exhibiting flat dispersion around the BIC\cite{RN719}(\autoref{fig:TM_band}(g)), as reflected in the featured far-field PL spectra (\autoref{fig:TM_band}(h)). In contrast, a grating with weak non-Hermiticity ($\Re[n_\text{low}]=3$) shows a higher $Q$ factor and intrinsic dispersion (\autoref{fig:TM_band}(c, d)). Notably, despite the large difference in non-Hermiticity between the two systems, the spatial field distributions of the BICs remain nearly identical (\autoref{fig:TM_band}(b, f)). This implies similar effective mode volumes $V_\text{eff}$, which isolates the $Q$ factor’s role in SE enhancement.

\begin{figure*}
    \centering
    \includegraphics[width=1\linewidth]{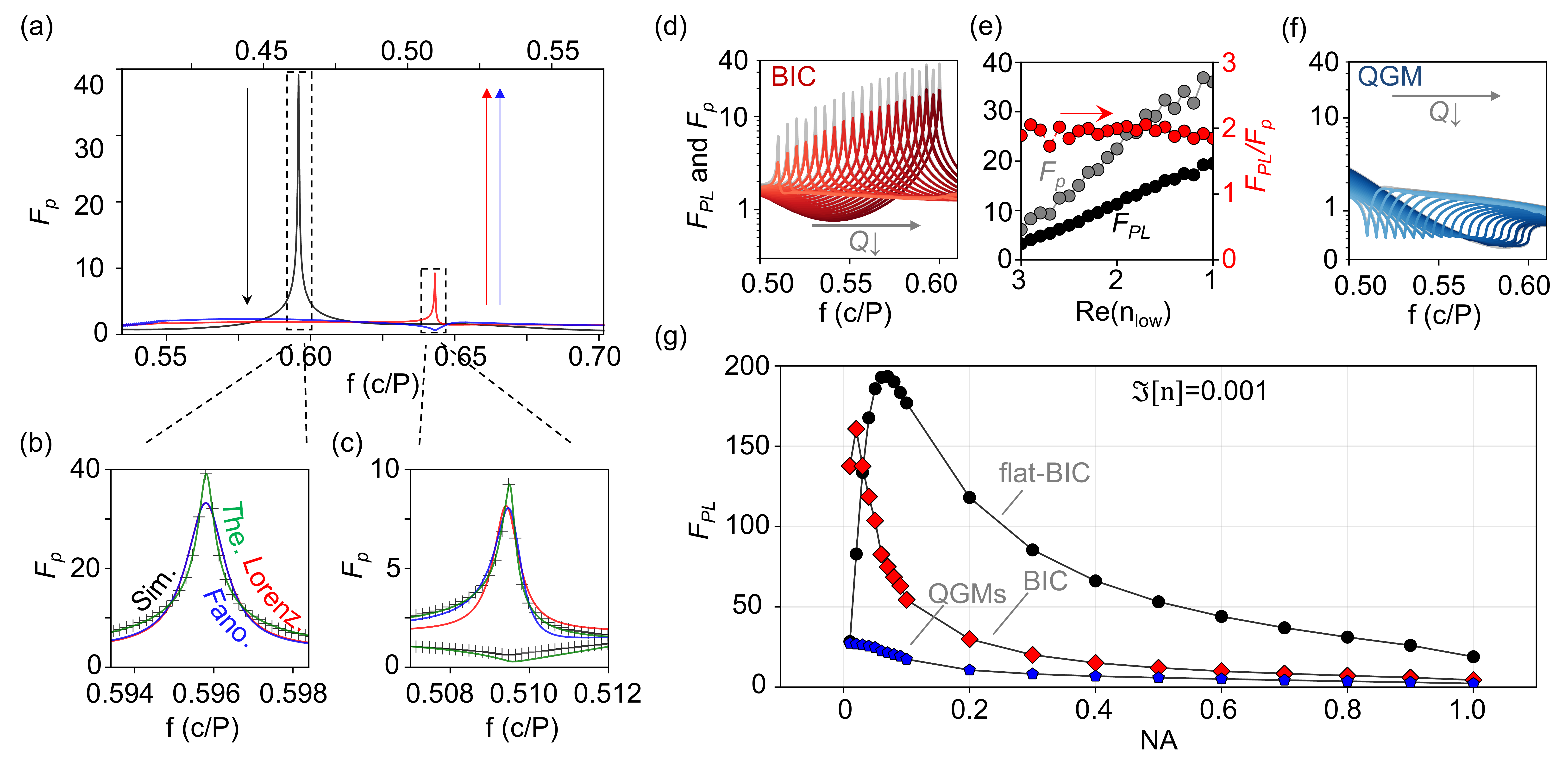}
    \caption{
    (a) Purcell factor for flat-BIC (black line), conventional BIC (red line), and QGM (blue line). (b,c) Analytical fit between numerical and theoretical spectrum, corresponding to the dashed boxes in (a), respectively. (d,f) Purcell (gray solid) and PL (colored solid) factors for BIC and QGM under varying low-index value, $\Re[n_\text{low}]$. (e) Peak values of $F_p$ and $F_\text{PL}$ from (d) versus $\Re[n_\text{low}]$. (g) PL enhancement under different NA collections. Black: flat-BIC. Red: conventional BIC. Blue: conventional QGM.
    }
    \label{fig:TM_SE}
\end{figure*}

We further evaluate the Purcell and PL factors using FDTD simulations of an array of over 200 grating periods. As shown in \autoref{fig:TM_SE}(a), the flat-BIC system (black line) exhibits significantly stronger SE enhancement than conventional high-$Q$ BICs (red) and QGMs (blue), achieving a peak Purcell factor of $F_p \approx 40$ near $f \approx 0.6$ with a non-Lorentzian, non-Fano spectral shape, as predicted. In \autoref{fig:TM_SE}(b), the simulated spectral features (cross markers) validate our theory, with Lorentzian (red) and Fano (blue) models failing to fit the simulation data, while our integrated theoretical model (green line) aligns closely. For QGM excitation, a spectral dip occurs near $f \approx 0.5$, consistent with predictions. In the high-$Q$ case(\autoref{fig:TM_SE}(c)), both BIC-excitation peak and QGM-excitation dip are quantitatively match our theoretical lineshape (green line). Such features could be directly observed through time-resolved fluorescence decay measurements\cite{RN1013}. Details of lineshape fitting are provided in the Supplementary Materials  Sec. 6.\cite{supp1}

By continuously tuning the grating’s refractive index contrast through adjustment of  $\Re[n_\text{low}]$, we transform the high-$Q$ grating into the low-$Q$ one, with corresponding spectra shown in \autoref{fig:TM_SE}(d, f). As the $Q$ factor decreases, SE enhancement under BIC excitation increases anomalously, in excellent agreement with our theory. Moreover, as shown in \autoref{fig:TM_SE}(e), $F_p$ remains consistently twice $F_{\rm PL}$, as predicted by Eq.~\eqref{eq:twice_relation_Fp_FPL}, confirming the absence of critical coupling and a radiative efficiency of approximately 0.5. Further details are provided in the Supplementary Materials  Sec. 6.\cite{supp1}

Finally, we elucidate the distinct SE enhancement characteristics of QGMs and BICs, offering insights for designing high-performance PL metasurfaces. QGMs radiate primarily at normal incidence, offering PL enhancement within a narrow angular cone, while flat-BICs radiate over a broader angular range. This difference is quantified in \autoref{fig:TM_SE}(g), where QGMs provide an advantage only for very small NA collection. This NA-dependent lineshape is predictable by our theoretical framework, as detailed in Supplementary Materials  Sec. 2 and 6.\cite{supp1} In contrast, flat-BICs deliver superior PL enhancement across a wide NA range, validating our framework. These results highlight how non-Hermiticity, combined with topological protection, under the framework of collective resonance, enables robust SE enhancement, ideal for on-demand practical nanophotonic applications.


In conclusion, building on our validated predictions, we present a unified framework for SE in periodic metasurfaces, seamlessly separating radiative and absorptive channels while equally emphasizing momentum-space integration. Our central finding is an anomalous SE enhancement driven by the flat-BIC: non-Hermiticity, combined with topological protection, induces flat dispersion, boosting emission despite reduced single-mode $Q$ factors, thus eliminating the need for critical coupling. This mechanism unveils the distinct spectral profiles observed and underscores that collective Bloch mode effects, rather than single-mode properties, govern SE in periodic systems. Experimental validation through time-resolved and angle-resolved PL measurements can directly probe these dynamics and radiation patterns\cite{RN1013,RN657}. Overall, our findings establish a novel design paradigm leveraging non-Hermiticity and topological protection to achieve robust, broadband emission enhancements beyond conventional high-$Q$ resonators, and they motivate further exploration of flat-band physics and non-Hermitian photonics in light–matter interaction\cite{RN978,RN965,RN146,RN389}.

\bibliography{ref}

\begin{thebibliography}{52}%
\makeatletter
\providecommand \@ifxundefined [1]{%
 \@ifx{#1\undefined}
}%
\providecommand \@ifnum [1]{%
 \ifnum #1\expandafter \@firstoftwo
 \else \expandafter \@secondoftwo
 \fi
}%
\providecommand \@ifx [1]{%
 \ifx #1\expandafter \@firstoftwo
 \else \expandafter \@secondoftwo
 \fi
}%
\providecommand \natexlab [1]{#1}%
\providecommand \enquote  [1]{``#1''}%
\providecommand \bibnamefont  [1]{#1}%
\providecommand \bibfnamefont [1]{#1}%
\providecommand \citenamefont [1]{#1}%
\providecommand \href@noop [0]{\@secondoftwo}%
\providecommand \href [0]{\begingroup \@sanitize@url \@href}%
\providecommand \@href[1]{\@@startlink{#1}\@@href}%
\providecommand \@@href[1]{\endgroup#1\@@endlink}%
\providecommand \@sanitize@url [0]{\catcode `\\12\catcode `\$12\catcode `\&12\catcode `\#12\catcode `\^12\catcode `\_12\catcode `\%12\relax}%
\providecommand \@@startlink[1]{}%
\providecommand \@@endlink[0]{}%
\providecommand \url  [0]{\begingroup\@sanitize@url \@url }%
\providecommand \@url [1]{\endgroup\@href {#1}{\urlprefix }}%
\providecommand \urlprefix  [0]{URL }%
\providecommand \Eprint [0]{\href }%
\providecommand \doibase [0]{http://dx.doi.org/}%
\providecommand \selectlanguage [0]{\@gobble}%
\providecommand \bibinfo  [0]{\@secondoftwo}%
\providecommand \bibfield  [0]{\@secondoftwo}%
\providecommand \translation [1]{[#1]}%
\providecommand \BibitemOpen [0]{}%
\providecommand \bibitemStop [0]{}%
\providecommand \bibitemNoStop [0]{.\EOS\space}%
\providecommand \EOS [0]{\spacefactor3000\relax}%
\providecommand \BibitemShut  [1]{\csname bibitem#1\endcsname}%
\let\auto@bib@innerbib\@empty
\bibitem [{\citenamefont {Purcell}(1946)}]{PhysRev.69.674}%
  \BibitemOpen
  \bibfield  {author} {\bibinfo {author} {\bibfnamefont {E.~M.}\ \bibnamefont {Purcell}},\ }\href {\doibase 10.1103/PhysRev.69.674} {\bibfield  {journal} {\bibinfo  {journal} {Phys. Rev.}\ }\textbf {\bibinfo {volume} {69}},\ \bibinfo {pages} {674} (\bibinfo {year} {1946})}\BibitemShut {NoStop}%
\bibitem [{\citenamefont {Yu}\ \emph {et~al.}(2025)\citenamefont {Yu}, \citenamefont {Yao}, \citenamefont {Qiu},\ and\ \citenamefont {Li}}]{RN968}%
  \BibitemOpen
  \bibfield  {author} {\bibinfo {author} {\bibfnamefont {J.}~\bibnamefont {Yu}}, \bibinfo {author} {\bibfnamefont {W.}~\bibnamefont {Yao}}, \bibinfo {author} {\bibfnamefont {M.}~\bibnamefont {Qiu}}, \ and\ \bibinfo {author} {\bibfnamefont {Q.}~\bibnamefont {Li}},\ }\href {\doibase 10.1038/s41377-025-01825-x} {\bibfield  {journal} {\bibinfo  {journal} {Light: Science {\&} Applications}\ }\textbf {\bibinfo {volume} {14}},\ \bibinfo {pages} {174} (\bibinfo {year} {2025})}\BibitemShut {NoStop}%
\bibitem [{\citenamefont {Wang}\ \emph {et~al.}(2021)\citenamefont {Wang}, \citenamefont {Yu}, \citenamefont {Wang}, \citenamefont {Zhang}, \citenamefont {Kuo}, \citenamefont {Xu},\ and\ \citenamefont {Wang}}]{RN1253}%
  \BibitemOpen
  \bibfield  {author} {\bibinfo {author} {\bibfnamefont {B.}~\bibnamefont {Wang}}, \bibinfo {author} {\bibfnamefont {P.}~\bibnamefont {Yu}}, \bibinfo {author} {\bibfnamefont {W.}~\bibnamefont {Wang}}, \bibinfo {author} {\bibfnamefont {X.}~\bibnamefont {Zhang}}, \bibinfo {author} {\bibfnamefont {H.-C.}\ \bibnamefont {Kuo}}, \bibinfo {author} {\bibfnamefont {H.}~\bibnamefont {Xu}}, \ and\ \bibinfo {author} {\bibfnamefont {Z.~M.}\ \bibnamefont {Wang}},\ }\href {\doibase https://doi.org/10.1002/adom.202001520} {\bibfield  {journal} {\bibinfo  {journal} {Advanced Optical Materials}\ }\textbf {\bibinfo {volume} {9}},\ \bibinfo {pages} {2001520} (\bibinfo {year} {2021})}\BibitemShut {NoStop}%
\bibitem [{\citenamefont {Lalanne}\ \emph {et~al.}(2018{\natexlab{a}})\citenamefont {Lalanne}, \citenamefont {Yan}, \citenamefont {Vynck}, \citenamefont {Sauvan},\ and\ \citenamefont {Hugonin}}]{RN1235}%
  \BibitemOpen
  \bibfield  {author} {\bibinfo {author} {\bibfnamefont {P.}~\bibnamefont {Lalanne}}, \bibinfo {author} {\bibfnamefont {W.}~\bibnamefont {Yan}}, \bibinfo {author} {\bibfnamefont {K.}~\bibnamefont {Vynck}}, \bibinfo {author} {\bibfnamefont {C.}~\bibnamefont {Sauvan}}, \ and\ \bibinfo {author} {\bibfnamefont {J.-P.}\ \bibnamefont {Hugonin}},\ }\href {\doibase https://doi.org/10.1002/lpor.201700113} {\bibfield  {journal} {\bibinfo  {journal} {Laser \& Photonics Reviews}\ }\textbf {\bibinfo {volume} {12}},\ \bibinfo {pages} {1700113} (\bibinfo {year} {2018}{\natexlab{a}})}\BibitemShut {NoStop}%
\bibitem [{\citenamefont {Kewes}\ \emph {et~al.}(2018)\citenamefont {Kewes}, \citenamefont {Binkowski}, \citenamefont {Burger}, \citenamefont {Zschiedrich},\ and\ \citenamefont {Benson}}]{RN1237}%
  \BibitemOpen
  \bibfield  {author} {\bibinfo {author} {\bibfnamefont {G.}~\bibnamefont {Kewes}}, \bibinfo {author} {\bibfnamefont {F.}~\bibnamefont {Binkowski}}, \bibinfo {author} {\bibfnamefont {S.}~\bibnamefont {Burger}}, \bibinfo {author} {\bibfnamefont {L.}~\bibnamefont {Zschiedrich}}, \ and\ \bibinfo {author} {\bibfnamefont {O.}~\bibnamefont {Benson}},\ }\href@noop {} {\bibfield  {journal} {\bibinfo  {journal} {ACS Photonics}\ }\textbf {\bibinfo {volume} {5}},\ \bibinfo {pages} {4089} (\bibinfo {year} {2018})}\BibitemShut {NoStop}%
\bibitem [{\citenamefont {Pellegrino}\ \emph {et~al.}(2020)\citenamefont {Pellegrino}, \citenamefont {Balestri}, \citenamefont {Granchi}, \citenamefont {Ciardi}, \citenamefont {Intonti}, \citenamefont {Pagliano}, \citenamefont {Silov}, \citenamefont {Otten}, \citenamefont {Wu}, \citenamefont {Vynck}, \citenamefont {Lalanne}, \citenamefont {Fiore},\ and\ \citenamefont {Gurioli}}]{RN1233}%
  \BibitemOpen
  \bibfield  {author} {\bibinfo {author} {\bibfnamefont {D.}~\bibnamefont {Pellegrino}}, \bibinfo {author} {\bibfnamefont {D.}~\bibnamefont {Balestri}}, \bibinfo {author} {\bibfnamefont {N.}~\bibnamefont {Granchi}}, \bibinfo {author} {\bibfnamefont {M.}~\bibnamefont {Ciardi}}, \bibinfo {author} {\bibfnamefont {F.}~\bibnamefont {Intonti}}, \bibinfo {author} {\bibfnamefont {F.}~\bibnamefont {Pagliano}}, \bibinfo {author} {\bibfnamefont {A.~Y.}\ \bibnamefont {Silov}}, \bibinfo {author} {\bibfnamefont {F.~W.}\ \bibnamefont {Otten}}, \bibinfo {author} {\bibfnamefont {T.}~\bibnamefont {Wu}}, \bibinfo {author} {\bibfnamefont {K.}~\bibnamefont {Vynck}}, \bibinfo {author} {\bibfnamefont {P.}~\bibnamefont {Lalanne}}, \bibinfo {author} {\bibfnamefont {A.}~\bibnamefont {Fiore}}, \ and\ \bibinfo {author} {\bibfnamefont {M.}~\bibnamefont {Gurioli}},\ }\href {https://www.ncbi.nlm.nih.gov/pubmed/32281836} {\bibfield  {journal} {\bibinfo  {journal} {Phys Rev Lett}\ }\textbf {\bibinfo {volume} {124}},\ \bibinfo {pages}
  {123902} (\bibinfo {year} {2020})}\BibitemShut {NoStop}%
\bibitem [{\citenamefont {Takata}\ \emph {et~al.}(2021)\citenamefont {Takata}, \citenamefont {Nozaki}, \citenamefont {Kuramochi}, \citenamefont {Matsuo}, \citenamefont {Takeda}, \citenamefont {Fujii}, \citenamefont {Kita}, \citenamefont {Shinya},\ and\ \citenamefont {Notomi}}]{RN937}%
  \BibitemOpen
  \bibfield  {author} {\bibinfo {author} {\bibfnamefont {K.}~\bibnamefont {Takata}}, \bibinfo {author} {\bibfnamefont {K.}~\bibnamefont {Nozaki}}, \bibinfo {author} {\bibfnamefont {E.}~\bibnamefont {Kuramochi}}, \bibinfo {author} {\bibfnamefont {S.}~\bibnamefont {Matsuo}}, \bibinfo {author} {\bibfnamefont {K.}~\bibnamefont {Takeda}}, \bibinfo {author} {\bibfnamefont {T.}~\bibnamefont {Fujii}}, \bibinfo {author} {\bibfnamefont {S.}~\bibnamefont {Kita}}, \bibinfo {author} {\bibfnamefont {A.}~\bibnamefont {Shinya}}, \ and\ \bibinfo {author} {\bibfnamefont {M.}~\bibnamefont {Notomi}},\ }\href {\doibase 10.1364/OPTICA.412596} {\bibfield  {journal} {\bibinfo  {journal} {Optica}\ }\textbf {\bibinfo {volume} {8}},\ \bibinfo {pages} {184} (\bibinfo {year} {2021})}\BibitemShut {NoStop}%
\bibitem [{\citenamefont {Noda}\ \emph {et~al.}(2007)\citenamefont {Noda}, \citenamefont {Fujita},\ and\ \citenamefont {Asano}}]{RN1255}%
  \BibitemOpen
  \bibfield  {author} {\bibinfo {author} {\bibfnamefont {S.}~\bibnamefont {Noda}}, \bibinfo {author} {\bibfnamefont {M.}~\bibnamefont {Fujita}}, \ and\ \bibinfo {author} {\bibfnamefont {T.}~\bibnamefont {Asano}},\ }\href {<Go to ISI>://WOS:000248905600011} {\bibfield  {journal} {\bibinfo  {journal} {Nature Photonics}\ }\textbf {\bibinfo {volume} {1}},\ \bibinfo {pages} {449} (\bibinfo {year} {2007})}\BibitemShut {NoStop}%
\bibitem [{\citenamefont {Joannopoulos}\ \emph {et~al.}(2008)\citenamefont {Joannopoulos}, \citenamefont {Johnson}, \citenamefont {Winn},\ and\ \citenamefont {Meade}}]{Joannopoulos2008PhotonicCM}%
  \BibitemOpen
  \bibfield  {author} {\bibinfo {author} {\bibfnamefont {J.~D.}\ \bibnamefont {Joannopoulos}}, \bibinfo {author} {\bibfnamefont {S.~G.}\ \bibnamefont {Johnson}}, \bibinfo {author} {\bibfnamefont {J.~N.}\ \bibnamefont {Winn}}, \ and\ \bibinfo {author} {\bibfnamefont {R.~D.}\ \bibnamefont {Meade}}\ }(\bibinfo {year} {2008})\BibitemShut {NoStop}%
\bibitem [{\citenamefont {Wang}\ \emph {et~al.}(2024)\citenamefont {Wang}, \citenamefont {Li}, \citenamefont {Zhao}, \citenamefont {Qian}, \citenamefont {Wang}, \citenamefont {Wang}, \citenamefont {Zhou}, \citenamefont {Han}, \citenamefont {Peng}, \citenamefont {Shi},\ and\ \citenamefont {Zi}}]{RN296}%
  \BibitemOpen
  \bibfield  {author} {\bibinfo {author} {\bibfnamefont {J.}~\bibnamefont {Wang}}, \bibinfo {author} {\bibfnamefont {P.}~\bibnamefont {Li}}, \bibinfo {author} {\bibfnamefont {X.}~\bibnamefont {Zhao}}, \bibinfo {author} {\bibfnamefont {Z.}~\bibnamefont {Qian}}, \bibinfo {author} {\bibfnamefont {X.}~\bibnamefont {Wang}}, \bibinfo {author} {\bibfnamefont {F.}~\bibnamefont {Wang}}, \bibinfo {author} {\bibfnamefont {X.}~\bibnamefont {Zhou}}, \bibinfo {author} {\bibfnamefont {D.}~\bibnamefont {Han}}, \bibinfo {author} {\bibfnamefont {C.}~\bibnamefont {Peng}}, \bibinfo {author} {\bibfnamefont {L.}~\bibnamefont {Shi}}, \ and\ \bibinfo {author} {\bibfnamefont {J.}~\bibnamefont {Zi}},\ }\href {https://www.researching.cn/articles/OJdf2741c1daea1ea5} {\bibfield  {journal} {\bibinfo  {journal} {Photonics Insights}\ }\textbf {\bibinfo {volume} {3}},\ \bibinfo {pages} {R01} (\bibinfo {year} {2024})}\BibitemShut {NoStop}%
\bibitem [{\citenamefont {Zhen}\ \emph {et~al.}(2014)\citenamefont {Zhen}, \citenamefont {Hsu}, \citenamefont {Lu}, \citenamefont {Stone},\ and\ \citenamefont {Soljacic}}]{RN475}%
  \BibitemOpen
  \bibfield  {author} {\bibinfo {author} {\bibfnamefont {B.}~\bibnamefont {Zhen}}, \bibinfo {author} {\bibfnamefont {C.~W.}\ \bibnamefont {Hsu}}, \bibinfo {author} {\bibfnamefont {L.}~\bibnamefont {Lu}}, \bibinfo {author} {\bibfnamefont {A.~D.}\ \bibnamefont {Stone}}, \ and\ \bibinfo {author} {\bibfnamefont {M.}~\bibnamefont {Soljacic}},\ }\href {https://www.ncbi.nlm.nih.gov/pubmed/25554906} {\bibfield  {journal} {\bibinfo  {journal} {Phys Rev Lett}\ }\textbf {\bibinfo {volume} {113}},\ \bibinfo {pages} {257401} (\bibinfo {year} {2014})}\BibitemShut {NoStop}%
\bibitem [{\citenamefont {Sun}\ \emph {et~al.}(2024{\natexlab{a}})\citenamefont {Sun}, \citenamefont {Wang},\ and\ \citenamefont {Han}}]{RN872}%
  \BibitemOpen
  \bibfield  {author} {\bibinfo {author} {\bibfnamefont {K.}~\bibnamefont {Sun}}, \bibinfo {author} {\bibfnamefont {W.}~\bibnamefont {Wang}}, \ and\ \bibinfo {author} {\bibfnamefont {Z.}~\bibnamefont {Han}},\ }\href {\doibase 10.1103/PhysRevB.109.085426} {\bibfield  {journal} {\bibinfo  {journal} {Phys. Rev. B}\ }\textbf {\bibinfo {volume} {109}},\ \bibinfo {pages} {085426} (\bibinfo {year} {2024}{\natexlab{a}})}\BibitemShut {NoStop}%
\bibitem [{\citenamefont {Sauvan}\ \emph {et~al.}(2013)\citenamefont {Sauvan}, \citenamefont {Hugonin}, \citenamefont {Maksymov},\ and\ \citenamefont {Lalanne}}]{RN1234}%
  \BibitemOpen
  \bibfield  {author} {\bibinfo {author} {\bibfnamefont {C.}~\bibnamefont {Sauvan}}, \bibinfo {author} {\bibfnamefont {J.~P.}\ \bibnamefont {Hugonin}}, \bibinfo {author} {\bibfnamefont {I.~S.}\ \bibnamefont {Maksymov}}, \ and\ \bibinfo {author} {\bibfnamefont {P.}~\bibnamefont {Lalanne}},\ }\href {https://www.ncbi.nlm.nih.gov/pubmed/25167528} {\bibfield  {journal} {\bibinfo  {journal} {Phys Rev Lett}\ }\textbf {\bibinfo {volume} {110}},\ \bibinfo {pages} {237401} (\bibinfo {year} {2013})}\BibitemShut {NoStop}%
\bibitem [{\citenamefont {Zhou}\ \emph {et~al.}(2024)\citenamefont {Zhou}, \citenamefont {Liu}, \citenamefont {Zhu}, \citenamefont {Gromyko}, \citenamefont {Qiu},\ and\ \citenamefont {Wu}}]{RN914}%
  \BibitemOpen
  \bibfield  {author} {\bibinfo {author} {\bibfnamefont {W.}~\bibnamefont {Zhou}}, \bibinfo {author} {\bibfnamefont {J.}~\bibnamefont {Liu}}, \bibinfo {author} {\bibfnamefont {J.}~\bibnamefont {Zhu}}, \bibinfo {author} {\bibfnamefont {D.}~\bibnamefont {Gromyko}}, \bibinfo {author} {\bibfnamefont {C.}~\bibnamefont {Qiu}}, \ and\ \bibinfo {author} {\bibfnamefont {L.}~\bibnamefont {Wu}},\ }\href {\doibase 10.1063/5.0191494} {\bibfield  {journal} {\bibinfo  {journal} {APL Quantum}\ }\textbf {\bibinfo {volume} {1}},\ \bibinfo {pages} {016110} (\bibinfo {year} {2024})}\BibitemShut {NoStop}%
\bibitem [{\citenamefont {Ginzburg}\ \emph {et~al.}(2017)\citenamefont {Ginzburg}, \citenamefont {Roth}, \citenamefont {Nasir}, \citenamefont {Segovia}, \citenamefont {Krasavin}, \citenamefont {Levitt}, \citenamefont {Hirvonen}, \citenamefont {Wells}, \citenamefont {Suhling}, \citenamefont {Richards}, \citenamefont {Podolskiy},\ and\ \citenamefont {Zayats}}]{RN1243}%
  \BibitemOpen
  \bibfield  {author} {\bibinfo {author} {\bibfnamefont {P.}~\bibnamefont {Ginzburg}}, \bibinfo {author} {\bibfnamefont {D.~J.}\ \bibnamefont {Roth}}, \bibinfo {author} {\bibfnamefont {M.~E.}\ \bibnamefont {Nasir}}, \bibinfo {author} {\bibfnamefont {P.}~\bibnamefont {Segovia}}, \bibinfo {author} {\bibfnamefont {A.~V.}\ \bibnamefont {Krasavin}}, \bibinfo {author} {\bibfnamefont {J.}~\bibnamefont {Levitt}}, \bibinfo {author} {\bibfnamefont {L.~M.}\ \bibnamefont {Hirvonen}}, \bibinfo {author} {\bibfnamefont {B.}~\bibnamefont {Wells}}, \bibinfo {author} {\bibfnamefont {K.}~\bibnamefont {Suhling}}, \bibinfo {author} {\bibfnamefont {D.}~\bibnamefont {Richards}}, \bibinfo {author} {\bibfnamefont {V.~A.}\ \bibnamefont {Podolskiy}}, \ and\ \bibinfo {author} {\bibfnamefont {A.~V.}\ \bibnamefont {Zayats}},\ }\href {https://www.ncbi.nlm.nih.gov/pubmed/30167260} {\bibfield  {journal} {\bibinfo  {journal} {Light Sci Appl}\ }\textbf {\bibinfo {volume} {6}},\ \bibinfo {pages} {e16273} (\bibinfo {year} {2017})}\BibitemShut
  {NoStop}%
\bibitem [{\citenamefont {Carminati}\ \emph {et~al.}(2015)\citenamefont {Carminati}, \citenamefont {Cazé}, \citenamefont {Cao}, \citenamefont {Peragut}, \citenamefont {Krachmalnicoff}, \citenamefont {Pierrat},\ and\ \citenamefont {De~Wilde}}]{RN1011}%
  \BibitemOpen
  \bibfield  {author} {\bibinfo {author} {\bibfnamefont {R.}~\bibnamefont {Carminati}}, \bibinfo {author} {\bibfnamefont {A.}~\bibnamefont {Cazé}}, \bibinfo {author} {\bibfnamefont {D.}~\bibnamefont {Cao}}, \bibinfo {author} {\bibfnamefont {F.}~\bibnamefont {Peragut}}, \bibinfo {author} {\bibfnamefont {V.}~\bibnamefont {Krachmalnicoff}}, \bibinfo {author} {\bibfnamefont {R.}~\bibnamefont {Pierrat}}, \ and\ \bibinfo {author} {\bibfnamefont {Y.}~\bibnamefont {De~Wilde}},\ }\href@noop {} {\bibfield  {journal} {\bibinfo  {journal} {Surface Science Reports}\ }\textbf {\bibinfo {volume} {70}},\ \bibinfo {pages} {1} (\bibinfo {year} {2015})}\BibitemShut {NoStop}%
\bibitem [{\citenamefont {Yuan}\ \emph {et~al.}(2017)\citenamefont {Yuan}, \citenamefont {Qiu}, \citenamefont {Cui}, \citenamefont {Zhu}, \citenamefont {Wang}, \citenamefont {Li}, \citenamefont {Song}, \citenamefont {Huang},\ and\ \citenamefont {Xia}}]{RN975}%
  \BibitemOpen
  \bibfield  {author} {\bibinfo {author} {\bibfnamefont {S.}~\bibnamefont {Yuan}}, \bibinfo {author} {\bibfnamefont {X.}~\bibnamefont {Qiu}}, \bibinfo {author} {\bibfnamefont {C.}~\bibnamefont {Cui}}, \bibinfo {author} {\bibfnamefont {L.}~\bibnamefont {Zhu}}, \bibinfo {author} {\bibfnamefont {Y.}~\bibnamefont {Wang}}, \bibinfo {author} {\bibfnamefont {Y.}~\bibnamefont {Li}}, \bibinfo {author} {\bibfnamefont {J.}~\bibnamefont {Song}}, \bibinfo {author} {\bibfnamefont {Q.}~\bibnamefont {Huang}}, \ and\ \bibinfo {author} {\bibfnamefont {J.}~\bibnamefont {Xia}},\ }\href@noop {} {\bibfield  {journal} {\bibinfo  {journal} {ACS Nano}\ }\textbf {\bibinfo {volume} {11}},\ \bibinfo {pages} {10704} (\bibinfo {year} {2017})}\BibitemShut {NoStop}%
\bibitem [{\citenamefont {Guo}\ \emph {et~al.}(2025)\citenamefont {Guo}, \citenamefont {Jin}, \citenamefont {Fu}, \citenamefont {Zhang}, \citenamefont {Yu}, \citenamefont {Chen}, \citenamefont {Wang}, \citenamefont {Huang}, \citenamefont {Zhou}, \citenamefont {Chen}, \citenamefont {Lu},\ and\ \citenamefont {Li}}]{RN967}%
  \BibitemOpen
  \bibfield  {author} {\bibinfo {author} {\bibfnamefont {J.}~\bibnamefont {Guo}}, \bibinfo {author} {\bibfnamefont {R.}~\bibnamefont {Jin}}, \bibinfo {author} {\bibfnamefont {Z.}~\bibnamefont {Fu}}, \bibinfo {author} {\bibfnamefont {Y.}~\bibnamefont {Zhang}}, \bibinfo {author} {\bibfnamefont {F.}~\bibnamefont {Yu}}, \bibinfo {author} {\bibfnamefont {J.}~\bibnamefont {Chen}}, \bibinfo {author} {\bibfnamefont {X.}~\bibnamefont {Wang}}, \bibinfo {author} {\bibfnamefont {L.}~\bibnamefont {Huang}}, \bibinfo {author} {\bibfnamefont {C.}~\bibnamefont {Zhou}}, \bibinfo {author} {\bibfnamefont {X.}~\bibnamefont {Chen}}, \bibinfo {author} {\bibfnamefont {W.}~\bibnamefont {Lu}}, \ and\ \bibinfo {author} {\bibfnamefont {G.}~\bibnamefont {Li}},\ }\href@noop {} {\bibfield  {journal} {\bibinfo  {journal} {Nano Letters}\ }\textbf {\bibinfo {volume} {25}},\ \bibinfo {pages} {2357} (\bibinfo {year} {2025})}\BibitemShut {NoStop}%
\bibitem [{\citenamefont {Wei}\ \emph {et~al.}(2025)\citenamefont {Wei}, \citenamefont {Wang}, \citenamefont {Daskalakis}, \citenamefont {Chou}, \citenamefont {Murai},\ and\ \citenamefont {Gomez~Rivas}}]{RN1247}%
  \BibitemOpen
  \bibfield  {author} {\bibinfo {author} {\bibfnamefont {Y.~C.}\ \bibnamefont {Wei}}, \bibinfo {author} {\bibfnamefont {C.~H.}\ \bibnamefont {Wang}}, \bibinfo {author} {\bibfnamefont {K.~S.}\ \bibnamefont {Daskalakis}}, \bibinfo {author} {\bibfnamefont {P.~T.}\ \bibnamefont {Chou}}, \bibinfo {author} {\bibfnamefont {S.}~\bibnamefont {Murai}}, \ and\ \bibinfo {author} {\bibfnamefont {J.}~\bibnamefont {Gomez~Rivas}},\ }\href {https://www.ncbi.nlm.nih.gov/pubmed/40255510} {\bibfield  {journal} {\bibinfo  {journal} {ACS Photonics}\ }\textbf {\bibinfo {volume} {12}},\ \bibinfo {pages} {2193} (\bibinfo {year} {2025})}\BibitemShut {NoStop}%
\bibitem [{\citenamefont {Zhang}\ \emph {et~al.}(2022)\citenamefont {Zhang}, \citenamefont {Liu}, \citenamefont {Han}, \citenamefont {Kivshar},\ and\ \citenamefont {Song}}]{RN988}%
  \BibitemOpen
  \bibfield  {author} {\bibinfo {author} {\bibfnamefont {X.}~\bibnamefont {Zhang}}, \bibinfo {author} {\bibfnamefont {Y.}~\bibnamefont {Liu}}, \bibinfo {author} {\bibfnamefont {J.}~\bibnamefont {Han}}, \bibinfo {author} {\bibfnamefont {Y.}~\bibnamefont {Kivshar}}, \ and\ \bibinfo {author} {\bibfnamefont {Q.}~\bibnamefont {Song}},\ }\href@noop {} {\bibfield  {journal} {\bibinfo  {journal} {Science}\ }\textbf {\bibinfo {volume} {377}},\ \bibinfo {pages} {1215} (\bibinfo {year} {2022})}\BibitemShut {NoStop}%
\bibitem [{\citenamefont {Liu}\ \emph {et~al.}(2018)\citenamefont {Liu}, \citenamefont {Vaskin}, \citenamefont {Addamane}, \citenamefont {Leung}, \citenamefont {Tsai}, \citenamefont {Yang}, \citenamefont {Vabishchevich}, \citenamefont {Keeler}, \citenamefont {Wang}, \citenamefont {He}, \citenamefont {Kim}, \citenamefont {Hartmann}, \citenamefont {Htoon}, \citenamefont {Doorn}, \citenamefont {Zilk}, \citenamefont {Pertsch}, \citenamefont {Balakrishnan}, \citenamefont {Sinclair}, \citenamefont {Staude},\ and\ \citenamefont {Brener}}]{RN1221}%
  \BibitemOpen
  \bibfield  {author} {\bibinfo {author} {\bibfnamefont {S.}~\bibnamefont {Liu}}, \bibinfo {author} {\bibfnamefont {A.}~\bibnamefont {Vaskin}}, \bibinfo {author} {\bibfnamefont {S.}~\bibnamefont {Addamane}}, \bibinfo {author} {\bibfnamefont {B.}~\bibnamefont {Leung}}, \bibinfo {author} {\bibfnamefont {M.-C.}\ \bibnamefont {Tsai}}, \bibinfo {author} {\bibfnamefont {Y.}~\bibnamefont {Yang}}, \bibinfo {author} {\bibfnamefont {P.~P.}\ \bibnamefont {Vabishchevich}}, \bibinfo {author} {\bibfnamefont {G.~A.}\ \bibnamefont {Keeler}}, \bibinfo {author} {\bibfnamefont {G.}~\bibnamefont {Wang}}, \bibinfo {author} {\bibfnamefont {X.}~\bibnamefont {He}}, \bibinfo {author} {\bibfnamefont {Y.}~\bibnamefont {Kim}}, \bibinfo {author} {\bibfnamefont {N.~F.}\ \bibnamefont {Hartmann}}, \bibinfo {author} {\bibfnamefont {H.}~\bibnamefont {Htoon}}, \bibinfo {author} {\bibfnamefont {S.~K.}\ \bibnamefont {Doorn}}, \bibinfo {author} {\bibfnamefont {M.}~\bibnamefont {Zilk}}, \bibinfo {author} {\bibfnamefont {T.}~\bibnamefont
  {Pertsch}}, \bibinfo {author} {\bibfnamefont {G.}~\bibnamefont {Balakrishnan}}, \bibinfo {author} {\bibfnamefont {M.~B.}\ \bibnamefont {Sinclair}}, \bibinfo {author} {\bibfnamefont {I.}~\bibnamefont {Staude}}, \ and\ \bibinfo {author} {\bibfnamefont {I.}~\bibnamefont {Brener}},\ }\href@noop {} {\bibfield  {journal} {\bibinfo  {journal} {Nano Letters}\ }\textbf {\bibinfo {volume} {18}},\ \bibinfo {pages} {6906} (\bibinfo {year} {2018})}\BibitemShut {NoStop}%
\bibitem [{\citenamefont {Kalinic}\ \emph {et~al.}(2023)\citenamefont {Kalinic}, \citenamefont {Cesca}, \citenamefont {Balasa}, \citenamefont {Trevisani}, \citenamefont {Jacassi}, \citenamefont {Maier}, \citenamefont {Sapienza},\ and\ \citenamefont {Mattei}}]{RN1246}%
  \BibitemOpen
  \bibfield  {author} {\bibinfo {author} {\bibfnamefont {B.}~\bibnamefont {Kalinic}}, \bibinfo {author} {\bibfnamefont {T.}~\bibnamefont {Cesca}}, \bibinfo {author} {\bibfnamefont {I.~G.}\ \bibnamefont {Balasa}}, \bibinfo {author} {\bibfnamefont {M.}~\bibnamefont {Trevisani}}, \bibinfo {author} {\bibfnamefont {A.}~\bibnamefont {Jacassi}}, \bibinfo {author} {\bibfnamefont {S.~A.}\ \bibnamefont {Maier}}, \bibinfo {author} {\bibfnamefont {R.}~\bibnamefont {Sapienza}}, \ and\ \bibinfo {author} {\bibfnamefont {G.}~\bibnamefont {Mattei}},\ }\href {https://www.ncbi.nlm.nih.gov/pubmed/36820324} {\bibfield  {journal} {\bibinfo  {journal} {ACS Photonics}\ }\textbf {\bibinfo {volume} {10}},\ \bibinfo {pages} {534} (\bibinfo {year} {2023})}\BibitemShut {NoStop}%
\bibitem [{\citenamefont {Zhang}\ \emph {et~al.}(2023)\citenamefont {Zhang}, \citenamefont {Xu}, \citenamefont {Liu}, \citenamefont {Lang}, \citenamefont {Zhang}, \citenamefont {Li}, \citenamefont {Lu}, \citenamefont {Chen}, \citenamefont {Wang}, \citenamefont {Wang},\ and\ \citenamefont {Li}}]{RN1229}%
  \BibitemOpen
  \bibfield  {author} {\bibinfo {author} {\bibfnamefont {Z.}~\bibnamefont {Zhang}}, \bibinfo {author} {\bibfnamefont {C.}~\bibnamefont {Xu}}, \bibinfo {author} {\bibfnamefont {C.}~\bibnamefont {Liu}}, \bibinfo {author} {\bibfnamefont {M.}~\bibnamefont {Lang}}, \bibinfo {author} {\bibfnamefont {Y.}~\bibnamefont {Zhang}}, \bibinfo {author} {\bibfnamefont {M.}~\bibnamefont {Li}}, \bibinfo {author} {\bibfnamefont {W.}~\bibnamefont {Lu}}, \bibinfo {author} {\bibfnamefont {Z.}~\bibnamefont {Chen}}, \bibinfo {author} {\bibfnamefont {C.}~\bibnamefont {Wang}}, \bibinfo {author} {\bibfnamefont {S.}~\bibnamefont {Wang}}, \ and\ \bibinfo {author} {\bibfnamefont {X.}~\bibnamefont {Li}},\ }\href {https://www.ncbi.nlm.nih.gov/pubmed/37539848} {\bibfield  {journal} {\bibinfo  {journal} {Nano Lett}\ }\textbf {\bibinfo {volume} {23}},\ \bibinfo {pages} {7584} (\bibinfo {year} {2023})}\BibitemShut {NoStop}%
\bibitem [{\citenamefont {Ren}\ \emph {et~al.}(2021)\citenamefont {Ren}, \citenamefont {Franke},\ and\ \citenamefont {Hughes}}]{RN995}%
  \BibitemOpen
  \bibfield  {author} {\bibinfo {author} {\bibfnamefont {J.}~\bibnamefont {Ren}}, \bibinfo {author} {\bibfnamefont {S.}~\bibnamefont {Franke}}, \ and\ \bibinfo {author} {\bibfnamefont {S.}~\bibnamefont {Hughes}},\ }\href {\doibase 10.1103/PhysRevX.11.041020} {\bibfield  {journal} {\bibinfo  {journal} {Phys. Rev. X}\ }\textbf {\bibinfo {volume} {11}},\ \bibinfo {pages} {041020} (\bibinfo {year} {2021})}\BibitemShut {NoStop}%
\bibitem [{\citenamefont {Franke}\ \emph {et~al.}(2021)\citenamefont {Franke}, \citenamefont {Ren}, \citenamefont {Richter}, \citenamefont {Knorr},\ and\ \citenamefont {Hughes}}]{RN1007}%
  \BibitemOpen
  \bibfield  {author} {\bibinfo {author} {\bibfnamefont {S.}~\bibnamefont {Franke}}, \bibinfo {author} {\bibfnamefont {J.}~\bibnamefont {Ren}}, \bibinfo {author} {\bibfnamefont {M.}~\bibnamefont {Richter}}, \bibinfo {author} {\bibfnamefont {A.}~\bibnamefont {Knorr}}, \ and\ \bibinfo {author} {\bibfnamefont {S.}~\bibnamefont {Hughes}},\ }\href {\doibase 10.1103/PhysRevLett.127.013602} {\bibfield  {journal} {\bibinfo  {journal} {Phys. Rev. Lett.}\ }\textbf {\bibinfo {volume} {127}},\ \bibinfo {pages} {013602} (\bibinfo {year} {2021})}\BibitemShut {NoStop}%
\bibitem [{\citenamefont {Seok}\ \emph {et~al.}(2011)\citenamefont {Seok}, \citenamefont {Jamshidi}, \citenamefont {Kim}, \citenamefont {Dhuey}, \citenamefont {Lakhani}, \citenamefont {Choo}, \citenamefont {Schuck}, \citenamefont {Cabrini}, \citenamefont {Schwartzberg}, \citenamefont {Bokor}, \citenamefont {Yablonovitch},\ and\ \citenamefont {Wu}}]{RN1004}%
  \BibitemOpen
  \bibfield  {author} {\bibinfo {author} {\bibfnamefont {T.~J.}\ \bibnamefont {Seok}}, \bibinfo {author} {\bibfnamefont {A.}~\bibnamefont {Jamshidi}}, \bibinfo {author} {\bibfnamefont {M.}~\bibnamefont {Kim}}, \bibinfo {author} {\bibfnamefont {S.}~\bibnamefont {Dhuey}}, \bibinfo {author} {\bibfnamefont {A.}~\bibnamefont {Lakhani}}, \bibinfo {author} {\bibfnamefont {H.}~\bibnamefont {Choo}}, \bibinfo {author} {\bibfnamefont {P.~J.}\ \bibnamefont {Schuck}}, \bibinfo {author} {\bibfnamefont {S.}~\bibnamefont {Cabrini}}, \bibinfo {author} {\bibfnamefont {A.~M.}\ \bibnamefont {Schwartzberg}}, \bibinfo {author} {\bibfnamefont {J.}~\bibnamefont {Bokor}}, \bibinfo {author} {\bibfnamefont {E.}~\bibnamefont {Yablonovitch}}, \ and\ \bibinfo {author} {\bibfnamefont {M.~C.}\ \bibnamefont {Wu}},\ }\href@noop {} {\bibfield  {journal} {\bibinfo  {journal} {Nano Letters}\ }\textbf {\bibinfo {volume} {11}},\ \bibinfo {pages} {2606} (\bibinfo {year} {2011})}\BibitemShut {NoStop}%
\bibitem [{\citenamefont {Han}\ \emph {et~al.}(2025)\citenamefont {Han}, \citenamefont {Lim}, \citenamefont {Lee}, \citenamefont {Kim},\ and\ \citenamefont {Jun}}]{RN1220}%
  \BibitemOpen
  \bibfield  {author} {\bibinfo {author} {\bibfnamefont {J.}~\bibnamefont {Han}}, \bibinfo {author} {\bibfnamefont {Y.}~\bibnamefont {Lim}}, \bibinfo {author} {\bibfnamefont {J.}~\bibnamefont {Lee}}, \bibinfo {author} {\bibfnamefont {S.}~\bibnamefont {Kim}}, \ and\ \bibinfo {author} {\bibfnamefont {Y.~C.}\ \bibnamefont {Jun}},\ }\href {\doibase https://doi.org/10.1002/lpor.202401923} {\bibfield  {journal} {\bibinfo  {journal} {Laser \& Photonics Reviews}\ }\textbf {\bibinfo {volume} {19}},\ \bibinfo {pages} {2401923} (\bibinfo {year} {2025})}\BibitemShut {NoStop}%
\bibitem [{\citenamefont {Schiattarella}\ \emph {et~al.}(2024)\citenamefont {Schiattarella}, \citenamefont {Romano}, \citenamefont {Sirleto}, \citenamefont {Mocella}, \citenamefont {Rendina}, \citenamefont {Lanzio}, \citenamefont {Riminucci}, \citenamefont {Schwartzberg}, \citenamefont {Cabrini}, \citenamefont {Chen}, \citenamefont {Liang}, \citenamefont {Liu},\ and\ \citenamefont {Zito}}]{RN746}%
  \BibitemOpen
  \bibfield  {author} {\bibinfo {author} {\bibfnamefont {C.}~\bibnamefont {Schiattarella}}, \bibinfo {author} {\bibfnamefont {S.}~\bibnamefont {Romano}}, \bibinfo {author} {\bibfnamefont {L.}~\bibnamefont {Sirleto}}, \bibinfo {author} {\bibfnamefont {V.}~\bibnamefont {Mocella}}, \bibinfo {author} {\bibfnamefont {I.}~\bibnamefont {Rendina}}, \bibinfo {author} {\bibfnamefont {V.}~\bibnamefont {Lanzio}}, \bibinfo {author} {\bibfnamefont {F.}~\bibnamefont {Riminucci}}, \bibinfo {author} {\bibfnamefont {A.}~\bibnamefont {Schwartzberg}}, \bibinfo {author} {\bibfnamefont {S.}~\bibnamefont {Cabrini}}, \bibinfo {author} {\bibfnamefont {J.}~\bibnamefont {Chen}}, \bibinfo {author} {\bibfnamefont {L.}~\bibnamefont {Liang}}, \bibinfo {author} {\bibfnamefont {X.}~\bibnamefont {Liu}}, \ and\ \bibinfo {author} {\bibfnamefont {G.}~\bibnamefont {Zito}},\ }\href {https://www.ncbi.nlm.nih.gov/pubmed/38383627} {\bibfield  {journal} {\bibinfo  {journal} {Nature}\ }\textbf {\bibinfo {volume} {626}},\ \bibinfo {pages} {765}
  (\bibinfo {year} {2024})}\BibitemShut {NoStop}%
\bibitem [{\citenamefont {Zhang}\ \emph {et~al.}(2015)\citenamefont {Zhang}, \citenamefont {Martins}, \citenamefont {Diyaf}, \citenamefont {Wilson}, \citenamefont {Turnbull},\ and\ \citenamefont {Samuel}}]{RN1223}%
  \BibitemOpen
  \bibfield  {author} {\bibinfo {author} {\bibfnamefont {S.}~\bibnamefont {Zhang}}, \bibinfo {author} {\bibfnamefont {E.~R.}\ \bibnamefont {Martins}}, \bibinfo {author} {\bibfnamefont {A.~G.}\ \bibnamefont {Diyaf}}, \bibinfo {author} {\bibfnamefont {J.~I.~B.}\ \bibnamefont {Wilson}}, \bibinfo {author} {\bibfnamefont {G.~A.}\ \bibnamefont {Turnbull}}, \ and\ \bibinfo {author} {\bibfnamefont {I.~D.~W.}\ \bibnamefont {Samuel}},\ }\href@noop {} {\bibfield  {journal} {\bibinfo  {journal} {Synthetic Metals}\ }\textbf {\bibinfo {volume} {205}},\ \bibinfo {pages} {127} (\bibinfo {year} {2015})}\BibitemShut {NoStop}%
\bibitem [{\citenamefont {Tse}\ \emph {et~al.}(2025{\natexlab{a}})\citenamefont {Tse}, \citenamefont {Enomoto}, \citenamefont {Murai},\ and\ \citenamefont {Tanaka}}]{RN1245}%
  \BibitemOpen
  \bibfield  {author} {\bibinfo {author} {\bibfnamefont {J.~T.~Y.}\ \bibnamefont {Tse}}, \bibinfo {author} {\bibfnamefont {T.}~\bibnamefont {Enomoto}}, \bibinfo {author} {\bibfnamefont {S.}~\bibnamefont {Murai}}, \ and\ \bibinfo {author} {\bibfnamefont {K.}~\bibnamefont {Tanaka}},\ }\href {https://arxiv.org/abs/2508.19710} {\enquote {\bibinfo {title} {Modelling purcell enhancement of metasurfaces supporting quasi-bound states in the continuum},}\ } (\bibinfo {year} {2025}{\natexlab{a}}),\ \Eprint {http://arxiv.org/abs/2508.19710} {arXiv:2508.19710 [physics.optics]} \BibitemShut {NoStop}%
\bibitem [{\citenamefont {Tse}\ \emph {et~al.}(2025{\natexlab{b}})\citenamefont {Tse}, \citenamefont {Murai},\ and\ \citenamefont {Tanaka}}]{RN1250}%
  \BibitemOpen
  \bibfield  {author} {\bibinfo {author} {\bibfnamefont {J.~T.~Y.}\ \bibnamefont {Tse}}, \bibinfo {author} {\bibfnamefont {S.}~\bibnamefont {Murai}}, \ and\ \bibinfo {author} {\bibfnamefont {K.}~\bibnamefont {Tanaka}},\ }\href {\doibase 10.1103/PhysRevA.111.013511} {\bibfield  {journal} {\bibinfo  {journal} {Phys. Rev. A}\ }\textbf {\bibinfo {volume} {111}},\ \bibinfo {pages} {013511} (\bibinfo {year} {2025}{\natexlab{b}})}\BibitemShut {NoStop}%
\bibitem [{\citenamefont {Maksimov}\ \emph {et~al.}(2025)\citenamefont {Maksimov}, \citenamefont {Pankin}, \citenamefont {Kim}, \citenamefont {Song}, \citenamefont {Peng},\ and\ \citenamefont {Bogdanov}}]{RN1249}%
  \BibitemOpen
  \bibfield  {author} {\bibinfo {author} {\bibfnamefont {D.~N.}\ \bibnamefont {Maksimov}}, \bibinfo {author} {\bibfnamefont {P.~S.}\ \bibnamefont {Pankin}}, \bibinfo {author} {\bibfnamefont {D.-W.}\ \bibnamefont {Kim}}, \bibinfo {author} {\bibfnamefont {M.}~\bibnamefont {Song}}, \bibinfo {author} {\bibfnamefont {C.}~\bibnamefont {Peng}}, \ and\ \bibinfo {author} {\bibfnamefont {A.~A.}\ \bibnamefont {Bogdanov}},\ }\href {https://arxiv.org/abs/2505.00396} {\enquote {\bibinfo {title} {Temporal coupled mode theory for high-q resonances in dielectric metasurfaces},}\ } (\bibinfo {year} {2025}),\ \Eprint {http://arxiv.org/abs/2505.00396} {arXiv:2505.00396 [physics.optics]} \BibitemShut {NoStop}%
\bibitem [{\citenamefont {Wonjoo}\ \emph {et~al.}(2004)\citenamefont {Wonjoo}, \citenamefont {Zheng},\ and\ \citenamefont {Shanhui}}]{RN239}%
  \BibitemOpen
  \bibfield  {author} {\bibinfo {author} {\bibfnamefont {S.}~\bibnamefont {Wonjoo}}, \bibinfo {author} {\bibfnamefont {W.}~\bibnamefont {Zheng}}, \ and\ \bibinfo {author} {\bibfnamefont {F.}~\bibnamefont {Shanhui}},\ }\href@noop {} {\bibfield  {journal} {\bibinfo  {journal} {IEEE Journal of Quantum Electronics}\ }\textbf {\bibinfo {volume} {40}},\ \bibinfo {pages} {1511} (\bibinfo {year} {2004})}\BibitemShut {NoStop}%
\bibitem [{\citenamefont {Gardiner}\ and\ \citenamefont {Collett}(1985)}]{PhysRevA.31.3761}%
  \BibitemOpen
  \bibfield  {author} {\bibinfo {author} {\bibfnamefont {C.~W.}\ \bibnamefont {Gardiner}}\ and\ \bibinfo {author} {\bibfnamefont {M.~J.}\ \bibnamefont {Collett}},\ }\href {\doibase 10.1103/PhysRevA.31.3761} {\bibfield  {journal} {\bibinfo  {journal} {Phys. Rev. A}\ }\textbf {\bibinfo {volume} {31}},\ \bibinfo {pages} {3761} (\bibinfo {year} {1985})}\BibitemShut {NoStop}%
\bibitem [{\citenamefont {VanDrunen}\ \emph {et~al.}(2024)\citenamefont {VanDrunen}, \citenamefont {Ren}, \citenamefont {Franke},\ and\ \citenamefont {Hughes}}]{RN1254}%
  \BibitemOpen
  \bibfield  {author} {\bibinfo {author} {\bibfnamefont {B.}~\bibnamefont {VanDrunen}}, \bibinfo {author} {\bibfnamefont {J.}~\bibnamefont {Ren}}, \bibinfo {author} {\bibfnamefont {S.}~\bibnamefont {Franke}}, \ and\ \bibinfo {author} {\bibfnamefont {S.}~\bibnamefont {Hughes}},\ }\href {\doibase 10.1364/OPTICAQ.504834} {\bibfield  {journal} {\bibinfo  {journal} {Optica Quantum}\ }\textbf {\bibinfo {volume} {2}},\ \bibinfo {pages} {85} (\bibinfo {year} {2024})}\BibitemShut {NoStop}%
\bibitem [{\citenamefont {Lalanne}\ \emph {et~al.}(2018{\natexlab{b}})\citenamefont {Lalanne}, \citenamefont {Coudert}, \citenamefont {Duchateau}, \citenamefont {Dilhaire},\ and\ \citenamefont {Vynck}}]{RN1277}%
  \BibitemOpen
  \bibfield  {author} {\bibinfo {author} {\bibfnamefont {P.}~\bibnamefont {Lalanne}}, \bibinfo {author} {\bibfnamefont {S.}~\bibnamefont {Coudert}}, \bibinfo {author} {\bibfnamefont {G.}~\bibnamefont {Duchateau}}, \bibinfo {author} {\bibfnamefont {S.}~\bibnamefont {Dilhaire}}, \ and\ \bibinfo {author} {\bibfnamefont {K.}~\bibnamefont {Vynck}},\ }\href {\doibase 10.1021/acsphotonics.8b01337} {\bibfield  {journal} {\bibinfo  {journal} {ACS Photonics}\ }\textbf {\bibinfo {volume} {6}},\ \bibinfo {pages} {4} (\bibinfo {year} {2018}{\natexlab{b}})}\BibitemShut {NoStop}%
\bibitem [{\citenamefont {Pick}\ \emph {et~al.}(2017)\citenamefont {Pick}, \citenamefont {Zhen}, \citenamefont {Miller}, \citenamefont {Hsu}, \citenamefont {Hernandez}, \citenamefont {Rodriguez}, \citenamefont {Soljacic},\ and\ \citenamefont {Johnson}}]{RN775}%
  \BibitemOpen
  \bibfield  {author} {\bibinfo {author} {\bibfnamefont {A.}~\bibnamefont {Pick}}, \bibinfo {author} {\bibfnamefont {B.}~\bibnamefont {Zhen}}, \bibinfo {author} {\bibfnamefont {O.~D.}\ \bibnamefont {Miller}}, \bibinfo {author} {\bibfnamefont {C.~W.}\ \bibnamefont {Hsu}}, \bibinfo {author} {\bibfnamefont {F.}~\bibnamefont {Hernandez}}, \bibinfo {author} {\bibfnamefont {A.~W.}\ \bibnamefont {Rodriguez}}, \bibinfo {author} {\bibfnamefont {M.}~\bibnamefont {Soljacic}}, \ and\ \bibinfo {author} {\bibfnamefont {S.~G.}\ \bibnamefont {Johnson}},\ }\href {https://www.ncbi.nlm.nih.gov/pubmed/28786590} {\bibfield  {journal} {\bibinfo  {journal} {Opt Express}\ }\textbf {\bibinfo {volume} {25}},\ \bibinfo {pages} {12325} (\bibinfo {year} {2017})}\BibitemShut {NoStop}%
\bibitem [{\citenamefont {Ferrier}\ \emph {et~al.}(2022)\citenamefont {Ferrier}, \citenamefont {Bouteyre}, \citenamefont {Pick}, \citenamefont {Cueff}, \citenamefont {Dang}, \citenamefont {Diederichs}, \citenamefont {Belarouci}, \citenamefont {Benyattou}, \citenamefont {Zhao}, \citenamefont {Su}, \citenamefont {Xing}, \citenamefont {Xiong},\ and\ \citenamefont {Nguyen}}]{RN913}%
  \BibitemOpen
  \bibfield  {author} {\bibinfo {author} {\bibfnamefont {L.}~\bibnamefont {Ferrier}}, \bibinfo {author} {\bibfnamefont {P.}~\bibnamefont {Bouteyre}}, \bibinfo {author} {\bibfnamefont {A.}~\bibnamefont {Pick}}, \bibinfo {author} {\bibfnamefont {S.}~\bibnamefont {Cueff}}, \bibinfo {author} {\bibfnamefont {N.~H.~M.}\ \bibnamefont {Dang}}, \bibinfo {author} {\bibfnamefont {C.}~\bibnamefont {Diederichs}}, \bibinfo {author} {\bibfnamefont {A.}~\bibnamefont {Belarouci}}, \bibinfo {author} {\bibfnamefont {T.}~\bibnamefont {Benyattou}}, \bibinfo {author} {\bibfnamefont {J.~X.}\ \bibnamefont {Zhao}}, \bibinfo {author} {\bibfnamefont {R.}~\bibnamefont {Su}}, \bibinfo {author} {\bibfnamefont {J.}~\bibnamefont {Xing}}, \bibinfo {author} {\bibfnamefont {Q.}~\bibnamefont {Xiong}}, \ and\ \bibinfo {author} {\bibfnamefont {H.~S.}\ \bibnamefont {Nguyen}},\ }\href {https://www.ncbi.nlm.nih.gov/pubmed/36053693} {\bibfield  {journal} {\bibinfo  {journal} {Phys Rev Lett}\ }\textbf {\bibinfo {volume} {129}},\ \bibinfo {pages}
  {083602} (\bibinfo {year} {2022})}\BibitemShut {NoStop}%
\bibitem [{\citenamefont {Lin}\ \emph {et~al.}(2016)\citenamefont {Lin}, \citenamefont {Pick}, \citenamefont {Loncar},\ and\ \citenamefont {Rodriguez}}]{RN709}%
  \BibitemOpen
  \bibfield  {author} {\bibinfo {author} {\bibfnamefont {Z.}~\bibnamefont {Lin}}, \bibinfo {author} {\bibfnamefont {A.}~\bibnamefont {Pick}}, \bibinfo {author} {\bibfnamefont {M.}~\bibnamefont {Loncar}}, \ and\ \bibinfo {author} {\bibfnamefont {A.~W.}\ \bibnamefont {Rodriguez}},\ }\href {https://www.ncbi.nlm.nih.gov/pubmed/27636493} {\bibfield  {journal} {\bibinfo  {journal} {Phys Rev Lett}\ }\textbf {\bibinfo {volume} {117}},\ \bibinfo {pages} {107402} (\bibinfo {year} {2016})}\BibitemShut {NoStop}%
\bibitem [{\citenamefont {Pelton}(2015)}]{RN1257}%
  \BibitemOpen
  \bibfield  {author} {\bibinfo {author} {\bibfnamefont {M.}~\bibnamefont {Pelton}},\ }\href@noop {} {\bibfield  {journal} {\bibinfo  {journal} {Nature Photonics}\ }\textbf {\bibinfo {volume} {9}},\ \bibinfo {pages} {427} (\bibinfo {year} {2015})}\BibitemShut {NoStop}%
\bibitem [{sup(2025)}]{supp1}%
  \BibitemOpen
  \href@noop {} {\enquote {\bibinfo {title} {Supplementary material: Derivations and simulation details},}\ }\bibinfo {howpublished} {Available online at \url{https://example.com/supp_material}} (\bibinfo {year} {2025}),\ \bibinfo {note} {this supplementary material includes derivations from TCMT to input-output formalism, Green's identity for power decomposition, EPs and BZF analyses, and full-wave simulation details with validation data.}\BibitemShut {Stop}%
\bibitem [{\citenamefont {Huang}\ \emph {et~al.}(2023)\citenamefont {Huang}, \citenamefont {Xu}, \citenamefont {Powell}, \citenamefont {Padilla},\ and\ \citenamefont {Miroshnichenko}}]{RN250}%
  \BibitemOpen
  \bibfield  {author} {\bibinfo {author} {\bibfnamefont {L.}~\bibnamefont {Huang}}, \bibinfo {author} {\bibfnamefont {L.}~\bibnamefont {Xu}}, \bibinfo {author} {\bibfnamefont {D.~A.}\ \bibnamefont {Powell}}, \bibinfo {author} {\bibfnamefont {W.~J.}\ \bibnamefont {Padilla}}, \ and\ \bibinfo {author} {\bibfnamefont {A.~E.}\ \bibnamefont {Miroshnichenko}},\ }\href@noop {} {\bibfield  {journal} {\bibinfo  {journal} {Physics Reports}\ }\textbf {\bibinfo {volume} {1008}},\ \bibinfo {pages} {1} (\bibinfo {year} {2023})}\BibitemShut {NoStop}%
\bibitem [{\citenamefont {Sun}\ \emph {et~al.}(2024{\natexlab{b}})\citenamefont {Sun}, \citenamefont {Levy},\ and\ \citenamefont {Han}}]{RN245}%
  \BibitemOpen
  \bibfield  {author} {\bibinfo {author} {\bibfnamefont {K.}~\bibnamefont {Sun}}, \bibinfo {author} {\bibfnamefont {U.}~\bibnamefont {Levy}}, \ and\ \bibinfo {author} {\bibfnamefont {Z.}~\bibnamefont {Han}},\ }\href {https://www.ncbi.nlm.nih.gov/pubmed/38166141} {\bibfield  {journal} {\bibinfo  {journal} {Nano Lett}\ }\textbf {\bibinfo {volume} {24}},\ \bibinfo {pages} {764} (\bibinfo {year} {2024}{\natexlab{b}})}\BibitemShut {NoStop}%
\bibitem [{\citenamefont {Koshelev}\ \emph {et~al.}(2018)\citenamefont {Koshelev}, \citenamefont {Lepeshov}, \citenamefont {Liu}, \citenamefont {Bogdanov},\ and\ \citenamefont {Kivshar}}]{RN220}%
  \BibitemOpen
  \bibfield  {author} {\bibinfo {author} {\bibfnamefont {K.}~\bibnamefont {Koshelev}}, \bibinfo {author} {\bibfnamefont {S.}~\bibnamefont {Lepeshov}}, \bibinfo {author} {\bibfnamefont {M.}~\bibnamefont {Liu}}, \bibinfo {author} {\bibfnamefont {A.}~\bibnamefont {Bogdanov}}, \ and\ \bibinfo {author} {\bibfnamefont {Y.}~\bibnamefont {Kivshar}},\ }\href {https://www.ncbi.nlm.nih.gov/pubmed/30468599} {\bibfield  {journal} {\bibinfo  {journal} {Phys Rev Lett}\ }\textbf {\bibinfo {volume} {121}},\ \bibinfo {pages} {193903} (\bibinfo {year} {2018})}\BibitemShut {NoStop}%
\bibitem [{\citenamefont {Zhen}\ \emph {et~al.}(2015)\citenamefont {Zhen}, \citenamefont {Hsu}, \citenamefont {Igarashi}, \citenamefont {Lu}, \citenamefont {Kaminer}, \citenamefont {Pick}, \citenamefont {Chua}, \citenamefont {Joannopoulos},\ and\ \citenamefont {Soljacic}}]{RN384}%
  \BibitemOpen
  \bibfield  {author} {\bibinfo {author} {\bibfnamefont {B.}~\bibnamefont {Zhen}}, \bibinfo {author} {\bibfnamefont {C.~W.}\ \bibnamefont {Hsu}}, \bibinfo {author} {\bibfnamefont {Y.}~\bibnamefont {Igarashi}}, \bibinfo {author} {\bibfnamefont {L.}~\bibnamefont {Lu}}, \bibinfo {author} {\bibfnamefont {I.}~\bibnamefont {Kaminer}}, \bibinfo {author} {\bibfnamefont {A.}~\bibnamefont {Pick}}, \bibinfo {author} {\bibfnamefont {S.~L.}\ \bibnamefont {Chua}}, \bibinfo {author} {\bibfnamefont {J.~D.}\ \bibnamefont {Joannopoulos}}, \ and\ \bibinfo {author} {\bibfnamefont {M.}~\bibnamefont {Soljacic}},\ }\href {https://www.ncbi.nlm.nih.gov/pubmed/26352476} {\bibfield  {journal} {\bibinfo  {journal} {Nature}\ }\textbf {\bibinfo {volume} {525}},\ \bibinfo {pages} {354} (\bibinfo {year} {2015})}\BibitemShut {NoStop}%
\bibitem [{\citenamefont {Wang}\ \emph {et~al.}(2025)\citenamefont {Wang}, \citenamefont {Sun}, \citenamefont {Ding}, \citenamefont {Zeng}, \citenamefont {Du}, \citenamefont {Han}, \citenamefont {Huang},\ and\ \citenamefont {Wang}}]{RN719}%
  \BibitemOpen
  \bibfield  {author} {\bibinfo {author} {\bibfnamefont {K.}~\bibnamefont {Wang}}, \bibinfo {author} {\bibfnamefont {K.}~\bibnamefont {Sun}}, \bibinfo {author} {\bibfnamefont {Q.}~\bibnamefont {Ding}}, \bibinfo {author} {\bibfnamefont {L.}~\bibnamefont {Zeng}}, \bibinfo {author} {\bibfnamefont {J.}~\bibnamefont {Du}}, \bibinfo {author} {\bibfnamefont {Z.}~\bibnamefont {Han}}, \bibinfo {author} {\bibfnamefont {L.}~\bibnamefont {Huang}}, \ and\ \bibinfo {author} {\bibfnamefont {W.}~\bibnamefont {Wang}},\ }\href {\doibase 10.1021/acs.nanolett.4c06565} {\bibfield  {journal} {\bibinfo  {journal} {Nano Letters}\ }\textbf {\bibinfo {volume} {25}},\ \bibinfo {pages} {3613} (\bibinfo {year} {2025})}\BibitemShut {NoStop}%
\bibitem [{\citenamefont {ter Huurne}\ \emph {et~al.}(2023)\citenamefont {ter Huurne}, \citenamefont {Peeters}, \citenamefont {Sánchez-Gil},\ and\ \citenamefont {Rivas}}]{RN1013}%
  \BibitemOpen
  \bibfield  {author} {\bibinfo {author} {\bibfnamefont {S.~E.~T.}\ \bibnamefont {ter Huurne}}, \bibinfo {author} {\bibfnamefont {D.~B.~L.}\ \bibnamefont {Peeters}}, \bibinfo {author} {\bibfnamefont {J.~A.}\ \bibnamefont {Sánchez-Gil}}, \ and\ \bibinfo {author} {\bibfnamefont {J.~G.}\ \bibnamefont {Rivas}},\ }\href@noop {} {\bibfield  {journal} {\bibinfo  {journal} {ACS Photonics}\ }\textbf {\bibinfo {volume} {10}},\ \bibinfo {pages} {2980} (\bibinfo {year} {2023})}\BibitemShut {NoStop}%
\bibitem [{\citenamefont {Zhang}\ \emph {et~al.}(2021)\citenamefont {Zhang}, \citenamefont {Zhao}, \citenamefont {Wang}, \citenamefont {Liu}, \citenamefont {Wang}, \citenamefont {Hu}, \citenamefont {Lu}, \citenamefont {Chen}, \citenamefont {Cui}, \citenamefont {Zhang}, \citenamefont {Hsu}, \citenamefont {Liu}, \citenamefont {Shi}, \citenamefont {Yin},\ and\ \citenamefont {Zi}}]{RN657}%
  \BibitemOpen
  \bibfield  {author} {\bibinfo {author} {\bibfnamefont {Y.}~\bibnamefont {Zhang}}, \bibinfo {author} {\bibfnamefont {M.}~\bibnamefont {Zhao}}, \bibinfo {author} {\bibfnamefont {J.}~\bibnamefont {Wang}}, \bibinfo {author} {\bibfnamefont {W.}~\bibnamefont {Liu}}, \bibinfo {author} {\bibfnamefont {B.}~\bibnamefont {Wang}}, \bibinfo {author} {\bibfnamefont {S.}~\bibnamefont {Hu}}, \bibinfo {author} {\bibfnamefont {G.}~\bibnamefont {Lu}}, \bibinfo {author} {\bibfnamefont {A.}~\bibnamefont {Chen}}, \bibinfo {author} {\bibfnamefont {J.}~\bibnamefont {Cui}}, \bibinfo {author} {\bibfnamefont {W.}~\bibnamefont {Zhang}}, \bibinfo {author} {\bibfnamefont {C.~W.}\ \bibnamefont {Hsu}}, \bibinfo {author} {\bibfnamefont {X.}~\bibnamefont {Liu}}, \bibinfo {author} {\bibfnamefont {L.}~\bibnamefont {Shi}}, \bibinfo {author} {\bibfnamefont {H.}~\bibnamefont {Yin}}, \ and\ \bibinfo {author} {\bibfnamefont {J.}~\bibnamefont {Zi}},\ }\href {https://www.ncbi.nlm.nih.gov/pubmed/36654139} {\bibfield  {journal} {\bibinfo  {journal}
  {Sci Bull (Beijing)}\ }\textbf {\bibinfo {volume} {66}},\ \bibinfo {pages} {824} (\bibinfo {year} {2021})}\BibitemShut {NoStop}%
\bibitem [{\citenamefont {Cui}\ \emph {et~al.}(2025)\citenamefont {Cui}, \citenamefont {Han}, \citenamefont {Zhu}, \citenamefont {Wang}, \citenamefont {Chua}, \citenamefont {Wang}, \citenamefont {Li}, \citenamefont {Davies}, \citenamefont {Linfield},\ and\ \citenamefont {Wang}}]{RN978}%
  \BibitemOpen
  \bibfield  {author} {\bibinfo {author} {\bibfnamefont {J.}~\bibnamefont {Cui}}, \bibinfo {author} {\bibfnamefont {S.}~\bibnamefont {Han}}, \bibinfo {author} {\bibfnamefont {B.}~\bibnamefont {Zhu}}, \bibinfo {author} {\bibfnamefont {C.}~\bibnamefont {Wang}}, \bibinfo {author} {\bibfnamefont {Y.}~\bibnamefont {Chua}}, \bibinfo {author} {\bibfnamefont {Q.}~\bibnamefont {Wang}}, \bibinfo {author} {\bibfnamefont {L.}~\bibnamefont {Li}}, \bibinfo {author} {\bibfnamefont {A.~G.}\ \bibnamefont {Davies}}, \bibinfo {author} {\bibfnamefont {E.~H.}\ \bibnamefont {Linfield}}, \ and\ \bibinfo {author} {\bibfnamefont {Q.~J.}\ \bibnamefont {Wang}},\ }\href {\doibase 10.1038/s41566-025-01665-6} {\bibfield  {journal} {\bibinfo  {journal} {Nature Photonics}\ }\textbf {\bibinfo {volume} {19}},\ \bibinfo {pages} {643} (\bibinfo {year} {2025})}\BibitemShut {NoStop}%
\bibitem [{\citenamefont {Sun}\ \emph {et~al.}(2025)\citenamefont {Sun}, \citenamefont {Wang}, \citenamefont {Wang}, \citenamefont {Cai}, \citenamefont {Huang}, \citenamefont {Alù},\ and\ \citenamefont {Han}}]{RN965}%
  \BibitemOpen
  \bibfield  {author} {\bibinfo {author} {\bibfnamefont {K.}~\bibnamefont {Sun}}, \bibinfo {author} {\bibfnamefont {K.}~\bibnamefont {Wang}}, \bibinfo {author} {\bibfnamefont {W.}~\bibnamefont {Wang}}, \bibinfo {author} {\bibfnamefont {Y.}~\bibnamefont {Cai}}, \bibinfo {author} {\bibfnamefont {L.}~\bibnamefont {Huang}}, \bibinfo {author} {\bibfnamefont {A.}~\bibnamefont {Alù}}, \ and\ \bibinfo {author} {\bibfnamefont {Z.}~\bibnamefont {Han}},\ }\href {\doibase https://doi.org/10.1016/j.newton.2025.100057} {\bibfield  {journal} {\bibinfo  {journal} {Newton}\ }\textbf {\bibinfo {volume} {1}},\ \bibinfo {pages} {100057} (\bibinfo {year} {2025})}\BibitemShut {NoStop}%
\bibitem [{\citenamefont {Le}\ \emph {et~al.}(2024)\citenamefont {Le}, \citenamefont {Bouteyre}, \citenamefont {Kheir-Aldine}, \citenamefont {Dubois}, \citenamefont {Cueff}, \citenamefont {Berguiga}, \citenamefont {Letartre}, \citenamefont {Viktorovitch}, \citenamefont {Benyattou},\ and\ \citenamefont {Nguyen}}]{RN146}%
  \BibitemOpen
  \bibfield  {author} {\bibinfo {author} {\bibfnamefont {N.~D.}\ \bibnamefont {Le}}, \bibinfo {author} {\bibfnamefont {P.}~\bibnamefont {Bouteyre}}, \bibinfo {author} {\bibfnamefont {A.}~\bibnamefont {Kheir-Aldine}}, \bibinfo {author} {\bibfnamefont {F.}~\bibnamefont {Dubois}}, \bibinfo {author} {\bibfnamefont {S.}~\bibnamefont {Cueff}}, \bibinfo {author} {\bibfnamefont {L.}~\bibnamefont {Berguiga}}, \bibinfo {author} {\bibfnamefont {X.}~\bibnamefont {Letartre}}, \bibinfo {author} {\bibfnamefont {P.}~\bibnamefont {Viktorovitch}}, \bibinfo {author} {\bibfnamefont {T.}~\bibnamefont {Benyattou}}, \ and\ \bibinfo {author} {\bibfnamefont {H.~S.}\ \bibnamefont {Nguyen}},\ }\href {https://www.ncbi.nlm.nih.gov/pubmed/38728718} {\bibfield  {journal} {\bibinfo  {journal} {Phys Rev Lett}\ }\textbf {\bibinfo {volume} {132}},\ \bibinfo {pages} {173802} (\bibinfo {year} {2024})}\BibitemShut {NoStop}%
\bibitem [{\citenamefont {Li}\ \emph {et~al.}(2023)\citenamefont {Li}, \citenamefont {Wei}, \citenamefont {Cotrufo}, \citenamefont {Chen}, \citenamefont {Mann}, \citenamefont {Ni}, \citenamefont {Xu}, \citenamefont {Chen}, \citenamefont {Wang}, \citenamefont {Fan}, \citenamefont {Qiu}, \citenamefont {Alu},\ and\ \citenamefont {Chen}}]{RN389}%
  \BibitemOpen
  \bibfield  {author} {\bibinfo {author} {\bibfnamefont {A.}~\bibnamefont {Li}}, \bibinfo {author} {\bibfnamefont {H.}~\bibnamefont {Wei}}, \bibinfo {author} {\bibfnamefont {M.}~\bibnamefont {Cotrufo}}, \bibinfo {author} {\bibfnamefont {W.}~\bibnamefont {Chen}}, \bibinfo {author} {\bibfnamefont {S.}~\bibnamefont {Mann}}, \bibinfo {author} {\bibfnamefont {X.}~\bibnamefont {Ni}}, \bibinfo {author} {\bibfnamefont {B.}~\bibnamefont {Xu}}, \bibinfo {author} {\bibfnamefont {J.}~\bibnamefont {Chen}}, \bibinfo {author} {\bibfnamefont {J.}~\bibnamefont {Wang}}, \bibinfo {author} {\bibfnamefont {S.}~\bibnamefont {Fan}}, \bibinfo {author} {\bibfnamefont {C.~W.}\ \bibnamefont {Qiu}}, \bibinfo {author} {\bibfnamefont {A.}~\bibnamefont {Alu}}, \ and\ \bibinfo {author} {\bibfnamefont {L.}~\bibnamefont {Chen}},\ }\href {https://www.ncbi.nlm.nih.gov/pubmed/37386141} {\bibfield  {journal} {\bibinfo  {journal} {Nat Nanotechnol}\ }\textbf {\bibinfo {volume} {18}},\ \bibinfo {pages} {706} (\bibinfo {year} {2023})}\BibitemShut
  {NoStop}%
\end{thebibliography}%
\bibliographystyle{apsrev4-1}

\end{document}